\documentclass[pdflatex,sn-mathphys-num]{sn-jnl}% Math and Physical Sciences Numbered Reference Style
\usepackage{graphicx}%
\usepackage{multirow}%
\usepackage{amsmath,amssymb,amsfonts}%
\usepackage{amsthm}%
\usepackage{mathrsfs}%
\usepackage[title]{appendix}%
\usepackage{xcolor}%
\usepackage{textcomp}%
\usepackage{manyfoot}%
\usepackage{booktabs}%
\usepackage{algorithm}%
\usepackage{algorithmicx}%
\usepackage{algpseudocode}%
\usepackage{listings}%
\usepackage{verbatim}
\usepackage{color}
\usepackage{lscape}
\usepackage[percent]{overpic}
\theoremstyle{thmstyleone}%
\theoremstyle{thmstyletwo}%

\theoremstyle{thmstylethree}%

\begin{document}

\title[Large Language Model Driven Development of Turbulence Models]{Large Language Model Driven Development of Turbulence Models}

%A path towards AI singularity in near-wall turbulence modeling via large language models

%%=============================================================%%
%% GivenName	-> \fnm{Joergen W.}
%% Particle	-> \spfx{van der} -> surname prefix
%% FamilyName	-> \sur{Ploeg}
%% Suffix	-> \sfx{IV}
%% \author*[1,2]{\fnm{Joergen W.} \spfx{van der} \sur{Ploeg} 
%%  \sfx{IV}}\email{iauthor@gmail.com}
%%=============================================================%%
\author[1]{\fnm{Zhongxin} \sur{Yang}}\email{zhongxinyang@stu.pku.edu}
\author[2,3]{\fnm{Yuanwei} \sur{Bin}}\email{ybin@eitech.edu}
\author[1]{\fnm{Yipeng} \sur{Shi}}\email{ypshi@coe.pku.edu}
\author*[4]{\fnm{Xiang I.A.} \sur{Yang}}\email{xzy48@psu.edu}

%\equalcont{These authors contributed equally to this work.}

\affil[1]{\orgdiv{College of Engineering}, \orgname{Peking University}, \orgaddress{\city{Beijing}, \postcode{100871}, \country{China}}}
\affil[2]{\orgdiv{Ningbo Institute of Digital Twin}, \orgname{Eastern Institute of Technology}, \orgaddress{\city{Ningbo}, \postcode{315200}, \state{Zhejiang}, \country{China}}}
\affil[3]{\orgname{Shenzhen Tenfong Technology Co., Ltd.}, \orgaddress{\city{Shenzhen}, \postcode{518000}, \state{Guangdong}, \country{China}}}
\affil*[4]{\orgdiv{Mechanical Engineering}, \orgname{Pennsylvania State University}, \orgaddress{\city{State College}, \postcode{16802}, \state{PA}, \country{USA}}}

%%==================================%%
%% Sample for unstructured abstract %%
%%==================================%%

\abstract{
Artificial intelligence (AI) has achieved human-level performance in specialized tasks such as Go, image recognition, and protein folding, raising the prospect of an AI singularity—where machines not only match but surpass human reasoning. Here, we demonstrate a step toward this vision in the context of turbulence modeling. By treating a large language model (LLM), DeepSeek-R1, as an equal partner, we establish a closed-loop, iterative workflow in which the LLM proposes, refines, and reasons about near-wall turbulence models under adverse pressure gradients (APGs), system rotation, and surface roughness. Through multiple rounds of interaction involving long-chain reasoning and a priori and a posteriori evaluations, the LLM generates models that not only rediscover established strategies but also synthesize new ones that outperform baseline wall models. Specifically, it recommends incorporating a material derivative to capture history effects in APG flows, modifying the law of the wall to account for system rotation, and developing rough-wall models informed by surface statistics. In contrast to conventional data-driven turbulence modeling—often characterized by human-designed, black-box architectures—the models developed here are physically interpretable and grounded in clear reasoning.
}

\keywords{AI singularity, Computational Modeling, Large Language Model, Computational Fluid Dynamics, Turbulence}

%%\pacs[JEL Classification]{D8, H51}

%%\pacs[MSC Classification]{35A01, 65L10, 65L12, 65L20, 65L70}

\maketitle

\section{Introduction}\label{sec:intro}

Artificial intelligence (AI) has achieved human-like performance in tasks once thought to require uniquely human intuition—mastering the game of Go \citep{silver2017mastering}, predicting protein structures \citep{jumper2021highly}, and driving through the streets \citep{yurtsever2020survey}. These advances have fueled speculation about the advent of General Artificial Intelligence (GAI), an AI that can reason, adapt, and solve problems across domains. A related concept gaining traction is the AI agent—a system capable of autonomous decision-making and iterative improvement in pursuit of a human-defined goal \citep{buehler2024mechgpt,ni2024mechagents,pandey2025openfoamgpt,dong2025fine}.
While most practical AI systems today remain narrow in scope, the emergence of large language models (LLMs) has narrowed the gap between domain-specific tools and general-purpose intelligence. These models, trained on vast corpora of human knowledge, can synthesize information, generate code, and reason over complex topics \citep{chang2024survey}.
Despite the hype surrounding GAI, compelling demonstrations of AI contributing new scientific insights—particularly in the physics—remain rare. 

Among the most enduring grand challenges in physics is turbulence—a chaotic, multi-scale phenomenon that resists closed-form description and predictive modeling. Despite over a century of effort, turbulence modeling remains largely empirical, guided by human intuition, physical reasoning, and hard-won insights from data \citep{meneveau2000scale,piomelli2002wall,durbin2018some}. Yet it underpins critical applications ranging from climate prediction \citep{alizadeh2022advances} and aerospace design \citep{mani2023perspective} to wind and energy systems \citep{stevens2017flow}. The gold standard of predictive fidelity, direct numerical simulation (DNS), is limited to canonical flows at modest Reynolds numbers due to its extreme computational cost \citep{yang2021grid,choi2012grid}. Large-eddy simulation (LES), which resolves large-scale motions while modeling smaller ones, offers a more tractable alternative \citep{goc2021large,goc2024wind}—but its cost remains prohibitive in high-Reynolds-number applications, especially near walls where turbulent eddies scale with their distance from the wall \citep{marusic2019attached}. This makes wall modeling the pacing item for extending LES to realistic flows \citep{bose2018wall,larsson2016large}.

\begin{figure}
    \centering
    \includegraphics[width=1.0\linewidth]{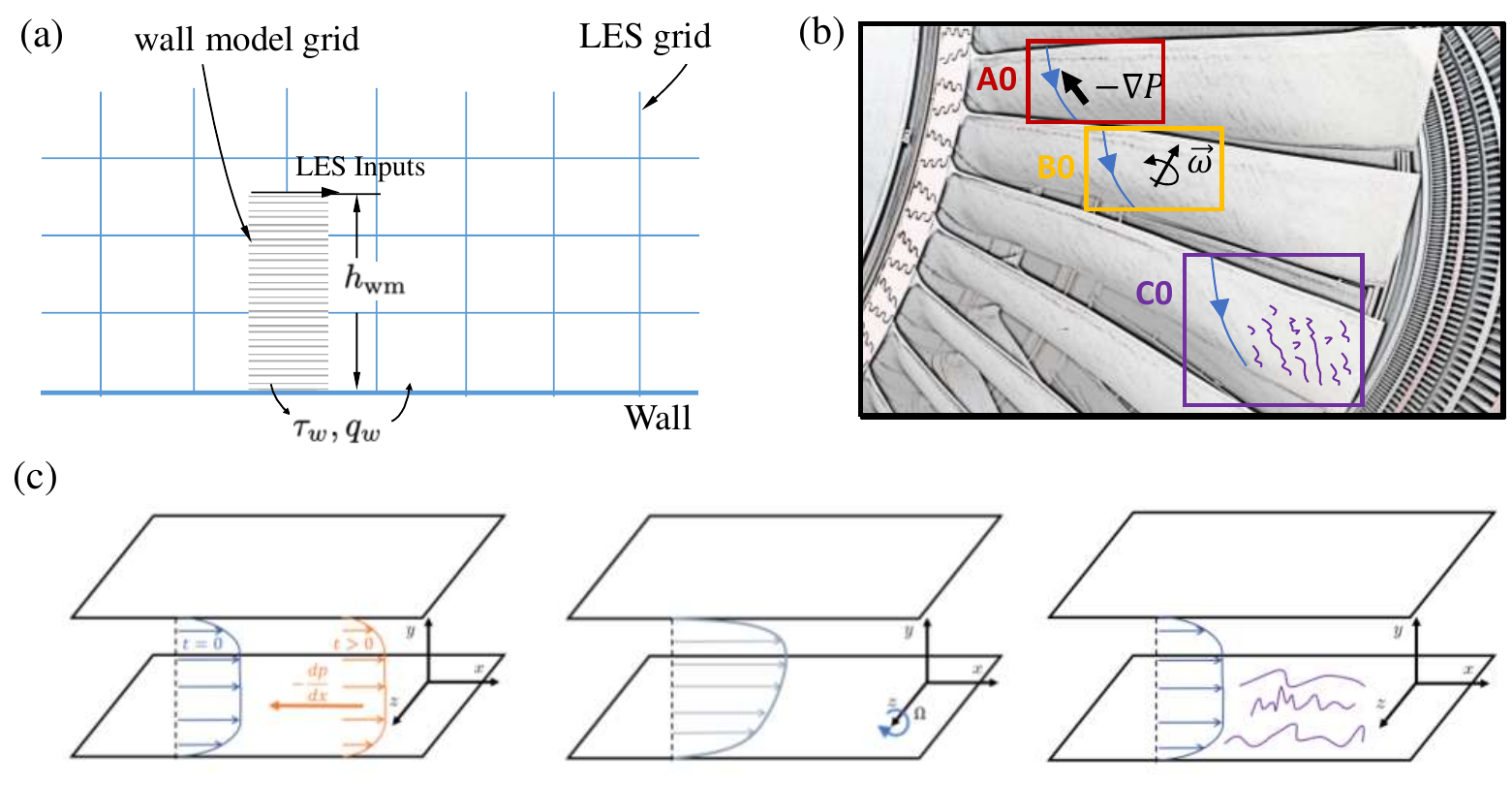}
    \caption{(a) Schematic of wall-modeled LES (WMLES). A wall model predicts the wall fluxes—shear stress $\tau_w$ and heat flux $q_w$—based on LES-resolved flow quantities at a distance $h_{\rm wm}$ from the wall. (b) Blades in a turbine, illustrating flows subjected non-equilibrium effects such as APGs ({\color{red} red box}), system rotation ({\color{orange} orange box}), and surface roughness ({\color{violet} purple box}). (c) Model problems. From left to righ: channel subjected a suddenly imposed APG, channel with system rotation, and channel with rough on the bottom wall.}
    \label{fig:Flowdemostration}
\end{figure}

Figure~\ref{fig:Flowdemostration}(a) schematically illustrates wall modeling in the context of LES. The LES grid in the near-wall region typically scales with the outer layer rather than the local eddies, leaving the wall layer unresolved. A wall model is therefore used to reconstruct the near-wall turbulence and predict wall fluxes, such as the wall shear stress $\tau_w$ and heat flux $q_w$, based on LES-resolved flow quantities at a distance $h_{\rm wm}$ from the wall—often referred to as the LES/wall-model matching location. The most widely used wall model is the equilibrium wall model (EWM) \citep{schumann1975subgrid,kawai2012wall,yang2017log}, which assumes the law of the wall (LoW) holds between the wall and $h_{\rm wm}$ locally and instantaneously. However, the LoW is valid only under equilibrium conditions for the mean flow.
Consequently, the EWM falls short when applied to the non-equilibrium boundary layers.
Figure~\ref{fig:Flowdemostration}(b) highlights a representative case in which near-wall turbulence is affected by non-equilibrium effects, including APGs, system rotation, and surface roughness. Efforts have been made to address these non-equilibrium effects in the contexts of wall models. In the following, we highlight a few of these developments. Park and Moin \cite{park2014improved} proposed a dynamic non-equilibrium wall model that accounts for the temporal lag between outer-layer changes and the wall shear stress response. Yang et al. \cite{yang2015integral} developed an integral wall model based on the momentum integral equation, enabling improved representation of mean velocity profiles. Bose and Moin \cite{bose2014dynamic} introduced a dynamic slip boundary condition that allows the wall shear stress to adapt to the large-scale structures resolved by LES. Building on this concept, Bae et al. \cite{bae2019dynamic} formulated a dynamic slip wall model with a self-consistent treatment of the slip length. Fowler et al. \cite{fowler2022lagrangian} introduced a Lagrangian relaxation wall model that does not impose the LoW instantaneously, but instead allows the modeled stress to evolve toward equilibrium over time, thereby better capturing non-equilibrium effects such as pressure gradients.
A more comprehensive review of recent progress in wall modeling is available in \cite{fowler2022lagrangian}, \cite{bose2018wall}, and \cite{yang2024predictive} and is not repeated here for brevity. While these advances have significantly expanded the applicability of LES, the process of model development remains human-driven—rooted in hypothesis generation, expert intuition, and iterative refinement. In this context, near-wall turbulence not only remains a persistent bottleneck but also serves as an ideal testbed for exploring whether AI can drive physical model development.

Recent years have seen a surge of interest in applying AI to turbulence modeling. In particular, machine learning (ML) tools have been used to develop turbulence closures by training on data from high-fidelity simulations and experiments \citep{duraisamy2019turbulence,pandey2020perspective,shan2023turbulence,bin2022progressive,bin2023data}. These efforts have yielded subgrid-scale models for LES \citep{maulik2019sub,cheng2022deep,xie2020modeling}, Reynolds stress closures for RANS \citep{ling2016reynolds,parish2016paradigm,wang2017physics,bin2024constrainedsa,bin2024constrainedsst,wu2025development}, and wall models for LES \citep{yang2019predictive,bae2022scientific,vadrot2023log,zhou2021wall,ma2025machine}.
Given the focus of this work on near-wall turbulence modeling, we briefly review several notable data-driven wall modeling efforts. Yang et al. \cite{yang2019predictive} proposed a predictive wall model based on supervised training of neural networks. The model learns a direct mapping between resolved flow quantities at the matching location and wall fluxes, with the training guided by the LoW. Zhou et al. \cite{zhou2021wall} adopted a similar supervised learning approach and demonstrated its effectiveness in LES of periodic hill flows, accurately capturing flow separation and reattachment.
\citet{bae2022scientific} introduced a multi-agent reinforcement learning (RL) framework, in which local agents infer wall stress through trial-and-error interactions with the LES environment, with rewards based on physical performance metrics such as velocity field accuracy. \citet{vadrot2023log} extended this RL approach and demonstrated that the trained model can recover the logarithmic law of the wall across a range of Reynolds numbers.
\citet{lozano2023machine} proposed a machine learning wall model based on the principle of composability. Rather than training on complex geometries directly, the model is trained on simple ``building-block” flows—canonical configurations such as turbulent channel and Couette flows—with the assumption that more complex boundary layer behaviors can be composed from these elementary patterns.
Additional reviews of recent developments in data-driven wall modeling can be found in \cite{vadrot2023survey} and are not repeated here for brevity. Despite these advances, the development process remains fundamentally human-guided. Researchers define the modeling objectives, curate training datasets, select architectures, and tune loss functions. The resulting models are typically evaluated on predefined benchmarks. 
This raises a broader question: can AI be tasked with open-ended problem solving in fluids engineering, where the objective is to iteratively generate and refine physically meaningful models?

In this study, we present a new paradigm in which a general-purpose LLM is tasked with solving problems in turbulence modeling. Unlike conventional applications where AI systems are trained to fit existing data within predefined architectures, we employ DeepSeek-R1, an open-weight LLM, as an autonomous agent operating within a closed-loop modeling framework. The LLM is prompted to generate strategies for near-wall turbulence modeling, which are then assessed both {\it a priori} and {\it a posteriori}. Based on performance feedback, the LLM revises its models through multiple iterations—mirroring the hypothesis generation, testing, and refinement cycles traditionally performed by human researchers. This approach departs from standard one-shot benchmark-driven machine learning workflows such as in \cite{prohl2024benchmarking} and \cite{jiang2025deepseek}. 
%Rather than optimizing fixed loss functions within static neural architectures, the LLM dynamically reformulates equations and modeling assumptions, guided by the outcomes of prior attempts. 
The present framework enables the model to engage in long-chain reasoning and to propose modeling strategies that exhibit both interpretability and performance.
We note that the choice of DeepSeek-R1 here over more widely used commercial models, such as ChatGPT or Claude, is motivated by two key considerations. First, DeepSeek-R1 is open-source, which enables full control over prompt design, reproducibility, and model deployment. Second, recent benchmarks suggest that DeepSeek-R1 exhibits strong reasoning capabilities, particularly in physics and mathematics domains \citep{gao2025comparison}.

This article focuses on three longstanding challenges in near-wall turbulence modeling: the effects of APGs, system rotation, and surface roughness. These physical mechanisms often appear simultaneously in engineering applications such as turbomachinery and aerospace flows. However, modeling their combined effects directly in complex configurations—such as those illustrated in Figure~\ref{fig:Flowdemostration}(b)—is ill-advised. The intertwined influence of multiple non-equilibrium effects makes it difficult to assess modeling errors. Here, we adopt a classical strategy in turbulence research: decomposing a complex modeling problem into a set of well-defined model problems. As sketched in Figure~\ref{fig:Flowdemostration}(c), we study each effect individually in the context of periodic channel flows, which provide controlled environments for isolating the impact of APG, rotation, or roughness. This serves multiple purposes. First, it enables scientific clarity—allowing us to assess whether the AI can identify the dominant mechanisms associated with each effect. Second, it facilitates validation against available high-fidelity data from DNS, which exists for these canonical flows—at least for mean velocity profiles and wall stresses \citep{yang2020mean,chen2023universal,huang2021bayesian,yang2023search,nair2024rough}. Third, it mirrors the historical development of turbulence models, where insight is gained from simplified configurations before generalization to more complex flow scenarios. For each model problem, DeepSeek-R1 is prompted to develop near-wall models. The proposed models are then evaluated in canonical LES settings, and their performance is used to iteratively refine subsequent formulations. In doing so, we shall see that the AI demonstrates an emerging capability for open-ended problem solving in fluids engineering.

The rest of the paper is organized as follows. 
In Section \ref{sec:Method}, the methodology is described.
Results are presented in Section \ref{sec:Results}, followed by concluding remarks in Section \ref{sec:Conclusion}.

\section{Methods}
\label{sec:Method}

\subsection{Large Language Model}
We use DeepSeek-R1 as our AI agent. DeepSeek-R1 is an LLM released in January 2025 by DeepSeek AI \citep{guo2025deepseek}. It is an open-weight, transformer-based model designed to support long-form reasoning.
%, with particular strength in mathematical and scientific domains. %\citep{deepseek2025eval}. 
The model is built on a dense decoder-only transformer architecture with 61 layers and approximately 67 billion parameters \citep{guo2025deepseek}. It supports a 32,000-token context window, enabling sustained long-chain reasoning across modeling iterations. The model was pretrained on a mixture of high-quality web documents, programming codes, and scientific texts, followed by instruction tuning to align its outputs with user intentions while preserving reasoning depth. 
%However, we note that the model was not fine-tuned on turbulence modeling tasks or closure-specific datasets.
The reasoning capabilities of Deepseek-R1 have been benchmarked against state-of-the-art models across scientific QA tasks, math word problems, and theorem proving, consistently showing competitive performance. 
%These features collectively support our goal of exploring autonomous model development using a general-purpose LLM with no turbulence-specific prior.
% \citep{deepseek2025eval}

When interacting with the LLM, we play the role of a fluids engineer, responsible for defining the modeling objectives and evaluating candidate solutions. The LLM, in turn, acts as the modeler, tasked with generating modeling strategies and outlining potential implementation pathways. The interaction begins by presenting the LLM with a flow configuration and the associated modeling challenge. It is then prompted to propose candidate solutions, often followed by clarification prompts to elicit additional details or justification needed for evaluation and practical implementation.
In contrast to benchmark-style evaluations, which typically involve a single round of interaction focused on verifying the correctness of LLM responses, our process seeks to mimic the communication between a turbulence modeler in academia and a fluids engineer in industry. This communication is inherently iterative, involving multiple rounds of feedback and refinement. Here, multiple rounds of interaction are critical: they provide opportunities for both the user and the LLM to correct omissions or misunderstandings. For example, the instructions we initially provide may be incomplete, or the models proposed by the LLM may contain errors. These shortcomings are often corrected through subsequent interactions, without explicit acknowledgment from either party.
%As we shall see in the subsequent sections, DeepSeek-R1 consistently exhibits a capacity for physical reasoning that extends beyond the mathematical structure of the Navier–Stokes equations. In particular, it provides physically grounded insights into boundary-layer behavior and turbulence mechanisms.
Once a potentially viable solution is provided, we proceed to formal evaluation. Two types of evaluation are employed. The first is based on feasibility: some solutions may require unavailable training data or demand impractical levels of manual coding, and such proposals are discarded. The second concerns accuracy, assessed through both a priori evaluations—focused on consistency with empirical knowledge—and a posteriori assessments via implementation and testing in CFD simulations. Models that do not demonstrate satisfactory agreement with DNS reference data are abandoned. Further details regarding the iterative interaction and evaluation process are provided in Section~\ref{sec:Results}.

\subsection{Computational Fluid Dynamics}

We solve the filtered incompressible Navier–Stokes equations, written in index notation with summation over repeated indices:
{\small\begin{equation} 
\partial_j \tilde{u}_j = 0, 
\label{eq:les_m} 
\end{equation} 
\begin{equation} 
\partial_t \tilde{u}_i + \partial_j \left( \tilde{u}_i \tilde{u}_j \right) = -\partial_i \left( \frac{\tilde{p}}{\rho} \right) + \nu \partial_j \partial_j \tilde{u}_i - \partial_j \tau_{ij}-2\epsilon_{ijk}\Omega_j \tilde{u}_k, 
\label{eq:les_u} 
\end{equation}}where $\tilde{u}_i$ and $\tilde{p}$ are the resolved velocity and pressure, respectively; $\rho$ is the fluid density, $\nu$ is the kinematic viscosity, and $\Omega_j$ denotes system rotation. The subgrid-scale (SGS) stress tensor $\tau_{ij}$ is modeled according to the Boussinesq hypothesis:
{\small\begin{equation} 
\tau_{ij} = -2\nu_t \tilde{S}_{ij} + \frac{1}{3} \tau_{kk} \delta_{ij}, 
\end{equation}}where $\tilde{S}_{ij} = \frac{1}{2} \left( \partial_j \tilde{u}_i + \partial_i \tilde{u}_j \right)$ is the resolved rate-of-strain tensor and $\delta_{ij}$ is the Kronecker delta. 
The eddy viscosity $\nu_t$ is computed using the Vreman model, whose details could be found in \cite{vreman2004eddy}.
Near-wall turbulence is not resolved and is modeled using a wall model. Two types are considered. The first is the EWM \citep{schumann1975subgrid,piomelli1989new,kawai2012wall,yang2017log}, which imposes the LoW between the wall and the LES/wall-model matching location $h_{\rm wm}$:
{\small\begin{equation} 
\tau_{w,x}/\rho=-\left[\frac{{U}^+_{||}}{f(y^+)}\right]^2\frac{\tilde{u}_{\rm LES}}{{U}_{||}},~ \tau_{w,z}/\rho=-\left[\frac{{U}^+_{||}}{f(y^+)}\right]^2\frac{\tilde{w}_{\rm LES}}{{ U}_{||}}
\label{eq:loglaw}
\end{equation}}where the superscript $+$ denotes normalization by wall units, ${\bf U}_{||}$ is the wall-parallel velocity, $\tilde{u}_{\rm LES}$ and $\tilde{w}_{\rm LES}$ are the LES velocity in the streamwise and the transverse directions at the LES/wall-model matching location.
In all of our WMLESs, $h_{wm}/h\approx 0.2$, and $f(y^+)$ is Spalding's LoW. 
The algebraic EWM in Eq. \eqref{eq:loglaw} has an ODE counterpart:
{\small\begin{equation}
    \frac{d}{dy}\left[(\nu+\nu_{t,{\rm wm}})\frac{d\tilde{u}_{||}}{dy}\right] = 0
    \label{eq:loglaw_ode}
\end{equation}}where
{\small\begin{equation}
\nu_t = \left[\kappa y (1-\exp(-y^+/A^+))\right]^2\Big|\frac{d\tilde{u}_{||}}{dy}\Big|, ~~A^+=26
\end{equation}}is the wall model eddy viscosity.
%Again, the models in \eqref{eq:loglaw} and \eqref{eq:loglaw_ode} are equivalent.   
%They 
The models in both Eqs. \eqref{eq:loglaw} and \eqref{eq:loglaw_ode} are equivalent.
They both impose the LoW between the matching location and the wall.
In the logarithmic region, the LoW is:
{\small\begin{equation} 
U^+ = \frac{1}{\kappa} \ln y^+ + B, 
\label{eq:log} 
\end{equation}}with von K{\'a}rm{\'a}n constant $\kappa \approx 0.41$ and intercept $B \approx 5.2$ \citep{pope2001turbulent}. For rough-wall flows, a roughness function $\Delta U^+$ is subtracted to reflect the downward shift in the velocity profile:
{\small\begin{equation} 
U^+ = \frac{1}{\kappa} \ln y^+ + B - \Delta U^+. 
\end{equation}}Besides the EWM, the second type of wall model includes those proposed from DeepSeek-R1. These LLM-driven closures are presented in Section~\ref{sec:Results}.
While the details of the models differ, their implementations are rather alike, and we place the LES/wall-model matching location at $h_{\rm wm}/h\approx 0.2$ as well.

\begin{table}
\caption{\label{tab:setup}Details of WMLESs. The friction Reynolds number is defined as $Re_\tau = u_\tau h / \nu$, where $u_\tau$ is the friction velocity, $h$ is the channel half-height, and $\nu$ is the kinematic viscosity. For cases involving a suddenly imposed APG, $Re_\tau$ is reported at the time instant when the APG is applied. The APG strength is characterized by $\Pi_0 = (h / \tau_{w,0}) (dp/dx)$, where $\tau_{w,0}$ is the wall shear stress prior to APG application. For cases with system rotation, the dimensionless rotation number is defined as $Ro_\tau = 2 h \Omega / u_\tau$, where $\Omega$ is the rotation rate. Surface roughness is parameterized using the non-dimensional equivalent sandgrain roughness height, $k_s^+ = k_s u_\tau / \nu$. Case labels use the abbreviations APG, ROT, and RW to indicate the presence of APG, rotation, and roughness, respectively. The prefix ``R" denotes the nominal Reynolds number and is followed by $Re_\tau / 100$.
}
{\small
\begin{tabular*}{\textwidth}{@{\extracolsep\fill}lcccc}
\toprule%
Case & $Re_\tau$ &  $L_x\times L_y \times L_z$ & $n_x \times n_y \times n_z$ & Remark \\
\midrule
& \multicolumn{4}{@{}c@{}}{APG channel} \\ \cmidrule{2-5}
R5APG1 & 544 & $2\pi h \times 2h \times 2\pi h$ & $64\times 64\times 64$ & APG channel, $\Pi_0 = 1$\\
R5APG10 & 544 & $2\pi h \times 2h \times 2\pi h$ & $64\times 64\times 64$ & APG channel, $\Pi_0 = 10$\\
R5APG100 & 544 & $2\pi h \times 2h \times 2\pi h$ & $64\times 64\times 64$ & APG channel, $\Pi_0 = 100$\\
R10APG10 & 1000 & $2\pi h \times 2h \times 2\pi h$ & $64\times 64\times 64$ & APG channel, $\Pi_0 = 10$\\ 
R10APG100 & 1000 & $2\pi h \times 2h \times 2\pi h$ & $64\times 64\times 64$ & APG channel, $\Pi_0 = 100$\\ \hline
& \multicolumn{4}{@{}c@{}}{Rotating channel} \\ \cmidrule{2-5}
R2ROT10 & 180 & $2\pi h \times 2h \times 2\pi h$ & $64\times 64\times 64$ & Rotating channel, $Ro_\tau = 10$\\
R2ROT22 & 180 & $2\pi h \times 2h \times 2\pi h$ & $64\times 64\times 64$ & Rotating channel, $Ro_\tau = 22$\\
R2ROT40 & 180 & $2\pi h \times 2h \times 2\pi h$ & $64\times 64\times 64$ & Rotating channel, $Ro_\tau = 40$\\
R2ROT80 & 180 & $2\pi h \times 2h \times 2\pi h$ & $64\times 64\times 64$ & Rotating channel, $Ro_\tau = 80$\\
R2ROT120 & 180 & $2\pi h \times 2h \times 2\pi h$ & $64\times 64\times 64$ & Rotating channel, $Ro_\tau = 120$\\
R4ROT10 & 360 & $2\pi h \times 2h \times 2\pi h$ & $64\times 64\times 64$ & Rotating channel, $Ro_\tau = 10$\\
R4ROT20 & 360 & $2\pi h \times 2h \times 2\pi h$ & $64\times 64\times 64$ & Rotating channel, $Ro_\tau = 20$\\
R4ROT32 & 360 & $2\pi h \times 2h \times 2\pi h$ & $64\times 64\times 64$ & Rotating channel, $Ro_\tau = 32$\\
R4ROT44 & 360 & $2\pi h \times 2h \times 2\pi h$ & $64\times 64\times 64$ & Rotating channel, $Ro_\tau = 44$\\ \hline
& \multicolumn{4}{@{}c@{}}{Rough wall channel} \\ \cmidrule{2-5}
R60RW1 & 6000 & $2\pi h \times 2h \times 2\pi h$ & $64\times 64\times 64$ & Sandgrain, $k_s^+ = 54.2$\\
R60RW2 & 6000 & $2\pi h \times 2h \times 2\pi h$ & $64\times 64\times 64$ & Sandgrain, $k_s^+ = 64.4$\\
R60RW3 & 6000 & $2\pi h \times 2h \times 2\pi h$ & $64\times 64\times 64$ & Grit-Blasted, $k_s^+ = 24.5$\\
R60RW4 & 6000 & $2\pi h \times 2h \times 2\pi h$ & $64\times 64\times 64$ & Grit-Blasted, $k_s^+ = 27.2$\\
R60RW5 & 6000 & $2\pi h \times 2h \times 2\pi h$ & $64\times 64\times 64$ & Grit-Blasted, $k_s^+ = 20.1$\\
R60RW6 & 6000 & $2\pi h \times 2h \times 2\pi h$ & $64\times 64\times 64$ & Grit-Blasted, $k_s^+ = 16.3$\\
R25RW7 & 2490 & $2\pi h \times 2h \times 2\pi h$ & $64\times 64\times 64$ & Truncated Core, $k_s^+ = 286$\\
R50RW8 & 4970 & $2\pi h \times 2h \times 2\pi h$ & $64\times 64\times 64$ & Multiscale LEGO-like, $k_s^+ = 646$\\
\botrule
\end{tabular*}
}
\end{table}

We focus on the model problems in figure \ref{fig:Flowdemostration}(c).
The flow configuration is a periodic channel. The flow is periodic in the streamwise and spanwise directions. The wall boundary condition is supplied by a wall model. 
%This configuration allows for isolated investigation of near-wall dynamics without the complexity of outer-layer boundary effects and has long been used for fundamental flow physics and modeling research.
%\citep{chung2009large,yang2020scaling,fowler2022lagrangian}.
Three flow effects are studied. First, fully developed plane channel subjected to a suddenly imposed APG. Second, plane channel subjected to spanwise system rotation. Third, channel flow with roughness on the bottom wall. 
%Each configuration isolates a key non-equilibrium mechanism relevant to wall-modeled LES (WMLES). 
Reference data are available in \cite{chen2023universal,xia2016direct,flack2016skin,medjnoun2021turbulent,womack2022turbulent,flack2023hydraulic}.
The computational grid is uniformly spaced in all directions following the standrad practice \citep{bose2014dynamic,park2014improved,yang2020scaling}. Key simulation parameters, including domain size, Reynolds number, and grid resolution, are listed in Table~\ref{tab:setup}. The domain size, grid spacing, and aspect ratios closely match those used in prior studies \citep{anderson2018turbulent,abkar2016minimum,martinez2018comparison,yang2024grid}, ensuring the credibility of performance evaluation and cross-study comparisons.

All simulations are performed using a finite-volume solver adapted for wall-modeled LES. The solver builds on the SUPES-cwm code base developed in \cite{lv2021wall} and \cite{gao2025novel}, which has been validated for channel flow calculations. It employs a fully conservative finite-volume formulation, nominal third-order spatial accuracy via characteristic-variable-based reconstruction, and an explicit strong-stability-preserving Runge–Kutta scheme for time integration. Numerical fluxes are computed using a hybrid scheme blending central differencing and HLLC fluxes, with dissipation controlled via a blending factor. Further details of the code's numerics could be found in  \cite{lv2021wall} and \cite{gao2025novel} along with a validation and are not repeated here for brevity.

\section{Results}
\label{sec:Results}

\subsection{Adverse Pressure Gradient}
\label{sec2_1:Case1}

The presence of a pressure gradient—whether favorable or adverse—modifies the velocity distribution and the shear stress within the boundary layer, leading to history effects \citep{bobke2017history,chen2023universal} and significant departures from the canonical LoW behavior observed under zero-pressure-gradient conditions \citep{perry2002streamwise,volino2020reynolds,pozuelo2022adverse}. This deviation complicates near-wall turbulence modeling, which often relies on equilibrium assumptions.
%Although empirical corrections and augmented wall models have been proposed \citep{kays2005convective,monkewitz2023hunt,knopp2022empirical}, these approaches often fail under strong or rapidly varying APGs due to their limited capacity to capture non-equilibrium and history effects \citep{bobke2017history,chen2023universal}. The difficulty stems from the non-local nature of the pressure gradient influence, which requires accounting for the cumulative history of wall shear development—an aspect not easily incorporated into conventional wall models.

\begin{figure}
    \centering
    \includegraphics[width=1\linewidth]{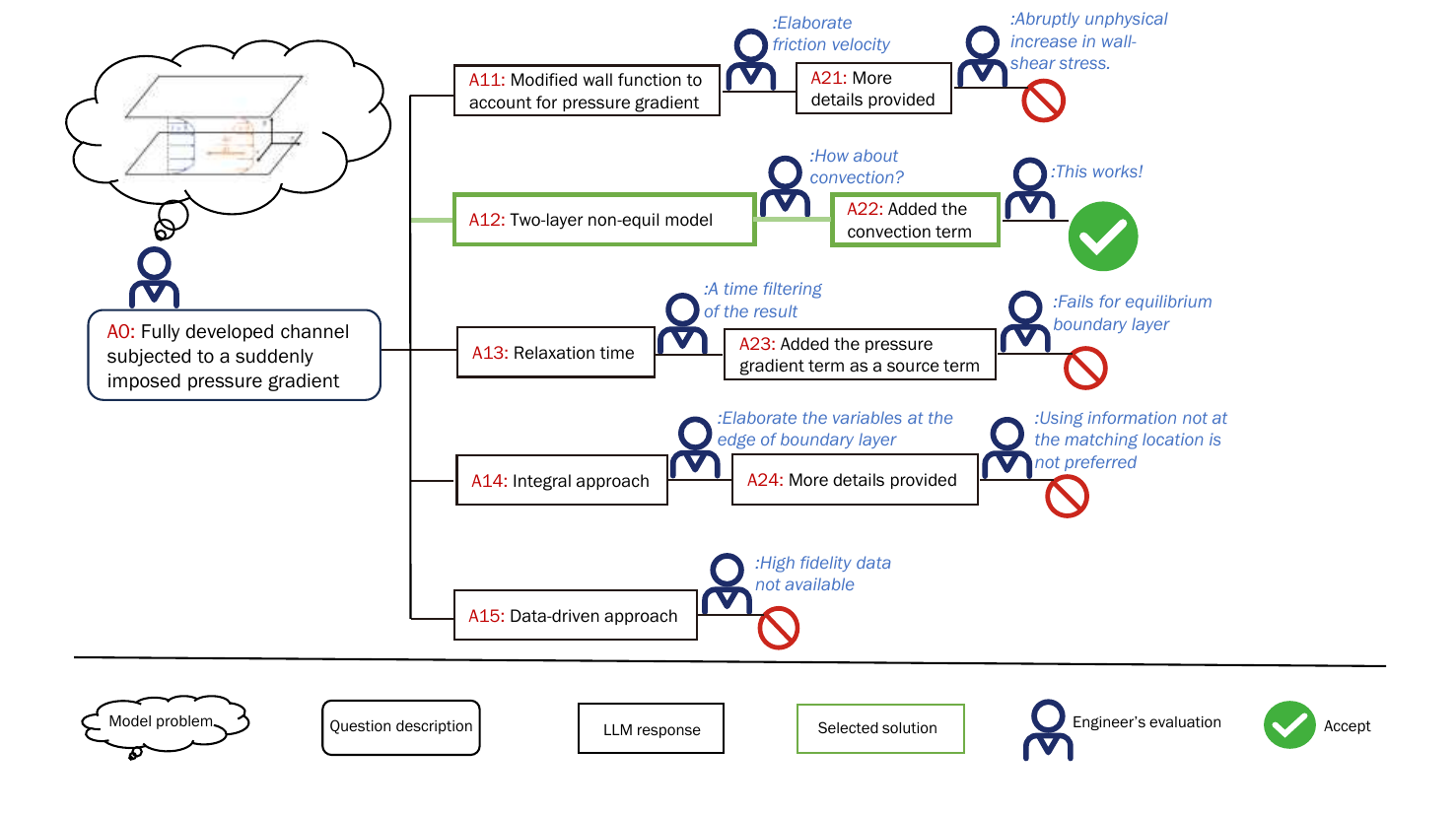}
    \caption{Schematic overview of the interaction between the LLM and the user for the APG modeling problem.}
    \label{fig:APG-sketch}
\end{figure}

In this subsection, we summarize our interactions with the LLM regarding this challenge, present the LLM-driven modeling solutions, and assess its performance. The modeling challenge was presented to the LLM using the following prompt:
\begin{quote}
\textit{I am a fluid engineer. I use wall-modeled large-eddy simulation. I have the following flow problem: a fully developed channel flow with a suddenly imposed APG. The imposed APG is kept constant. You can imagine that the friction on the wall will decrease. The logarithmic law based wall model is not effective in simulating this flow. Please propose new models, taking into account simplicity and generality.}
\end{quote}
An overview of the interaction between the LLM and the user is illustrated in Figure~\ref{fig:APG-sketch}, with the initial prompt summarized and labeled as A0. In the following, we document the communication process, highlighting how the LLM refines its reasoning, iteratively breaks down the problem, and ultimately converges on a viable solution.
The communication here uses about 7000 tokens.

The LLM proposes five potential modeling strategies, labeled A11 through A15 in figure \ref{fig:APG-sketch}. Here, the four less successful strategies are discussed first, followed by the more promising one.
The first strategy (A11) modifies the wall function. The LLM introduces the concept of an adjusted friction velocity, which we prompt it to elaborate on. The model takes the following form:
{\small\begin{equation}
\tau_w = \tau_{w,0} \left[ 1 + \alpha dP^+\right],
\label{eq:A11}
\end{equation}}where $\tau_{w,0}$ is the prediction of the EWM, $dP^+$ is the dimensionless pressure gradient, and $\alpha$ is a model constant. Although the intermediate model forms are not shown here for brevity, we note that the initial model form was dimensionally inconsistent. Nonetheless, this error was corrected in subsequent responses without explicit prompting—an encouraging behavior akin to human model development. Upon implementation, we observe an abrupt and unphysical change in wall shear stress upon the application of the APG, regardless of the choice of $\alpha$. Due to this fundamental misrepresentation of the flow physics, this strategy is discarded.
The third strategy (A13) introduces a relaxation timescale to the wall stress evolution:
{\small\begin{equation}
\tau_w^{n+1} = \tau_w^{n} + \Delta t \frac{\tau_{w,0} - \tau_w^{n}}{T_r},
\label{eq:A13}
\end{equation}}where $\tau_w^n$ and $\tau_w^{n+1}$ denote the wall shear stress at time steps $n$ and $n+1$, $\Delta t$ is the time step size, $\tau_{w,0}$ is the EWM prediction, and $T_r$ is a relaxation timescale. Analysis shows that this model acts as a low-pass temporal filter for wall shear stress but does not effectively capture pressure gradient effects. When this limitation is communicated to the LLM, it reformulates the model by adding a source term accounting for the APG:
{\small\begin{equation}
\partial_t \tau_w = \frac{\tau_{w,0} - \tau_w}{T_r} + \gamma u_\tau \partial_x P,
\end{equation}}where $\gamma$ is a model constant. {\it A posteriori} tests show that the model suffers from the same issue as the one in Eq. \eqref{eq:A11}, and is therefore discarded as well.
The fourth strategy (A14) is an integral model:
{\small\begin{equation}
\partial_t \theta + U_e\partial_x \theta = \frac{\tau_w}{\rho U_e^2} - \frac{\theta}{U_e}\partial_x U_e,
\label{eq:A14}
\end{equation}}Upon requesting clarification of the variables in the equation, we find that the model requires information at the edge of the boundary layer, which is generally not available in WMLES and is not desirable for practical implementation. 
%Additionally, the model in Eq.~\eqref{eq:A14} is dimensionally inconsistent. As we elected not to pursue this strategy further, the LLM was not given an opportunity to refine the model.
The fifth strategy (A15) suggests a data-driven modeling approach. However, the lack of high-fidelity training data renders this strategy infeasible within the scope of the current study.
The second strategy (A12) ultimately yields a viable solution. Upon follow-up discussions regarding the treatment of convective terms, the LLM refines the model from A12 to A22. The model is based on the thin boundary layer approximation, thus assuming $\partial_y P = 0$ and neglecting the wall-normal velocity. The final model takes the following form:
{\small\begin{equation}
\begin{split}
\frac{{\rm D} u_{\rm wm}}{{\rm D}t} = -\frac{1}{\rho}\frac{\partial  P_{\rm LES} }{\partial x}+ \frac{\partial}{\partial y} \left[ (\nu +\nu_{t,0})\frac{\partial u_{\rm wm}}{\partial y} \right],\\
\frac{{\rm D} w_{\rm wm}}{{\rm D}t} = -\frac{1}{\rho}\frac{\partial  P_{\rm LES} }{\partial z}+ \frac{\partial}{\partial y} \left[ (\nu +\nu_{t,0})\frac{\partial w_{\rm wm}}{\partial y} \right],
\end{split}
\label{eq:A22}
\end{equation}}Here, $u_{\rm wm}$ and $w_{\rm wm}$ are the velocity components in the $x$ and $z$ directions, respectively; $\nu$ is the molecular viscosity, and $\nu_{t,0}$ is the eddy viscosity, modeled in the same way as in the EWM. The model solves the velocity profiles between the LES/WM matching location and the wall in both wall-parallel directions, $x$ and $z$, according to Eq. \eqref{eq:A22}. The pressure gradients in both the $x$ and $z$ directions are obtained directly from the LES. Note that the pressure gradient and the material derivative terms require no closure, although storing the velocity profiles is necessary to compute the unsteady term. The rationale here is that history effects are largely captured by the material derivative and the pressure gradient, and no further modeling is required.
In addition to the reasoning and the model formulation, the LLM also supplies an initial Python implementation, which—although not directly portable to our CFD code—provides a useful starting point.

\begin{figure}
    \centering
    \includegraphics[width=0.32\linewidth]{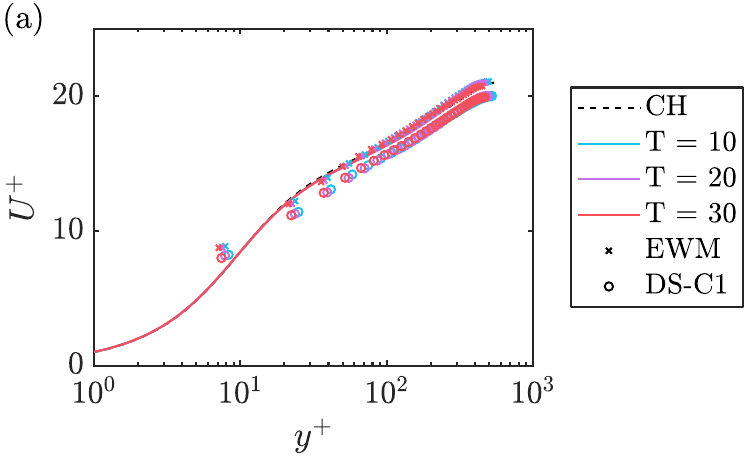}\includegraphics[width=0.32\linewidth]{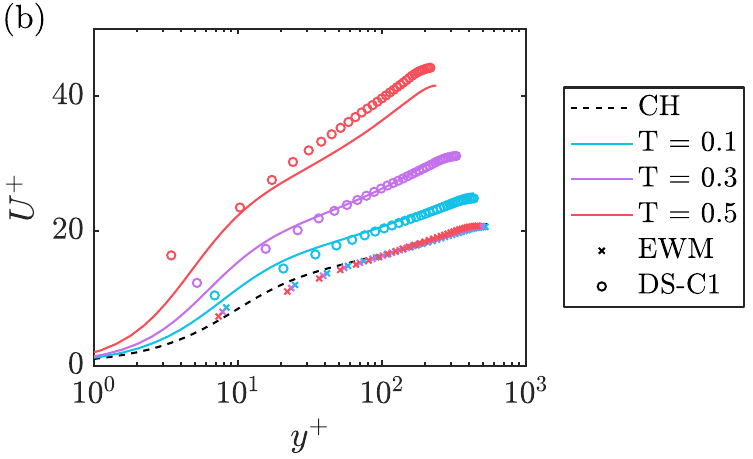}\includegraphics[width=0.32\linewidth]{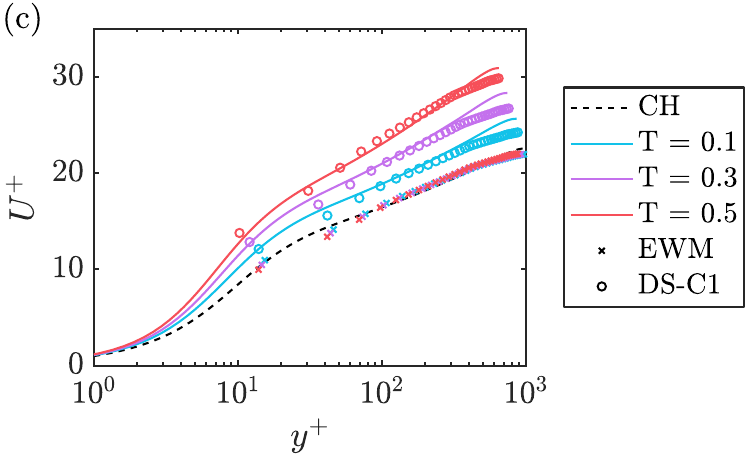}
    \includegraphics[width=0.32\linewidth]{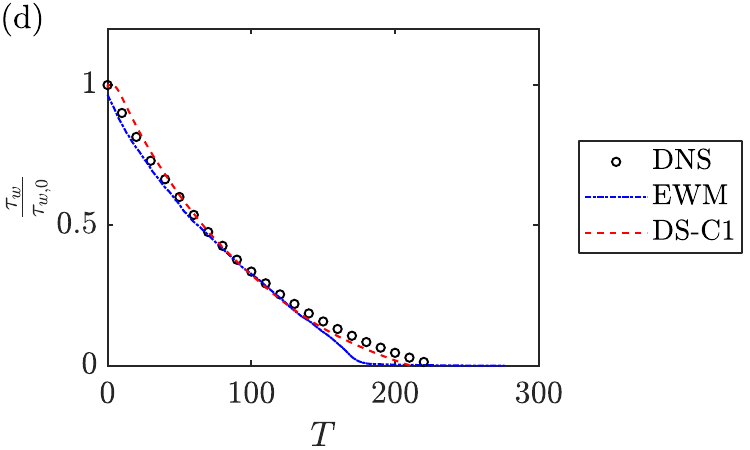}\includegraphics[width=0.32\linewidth]{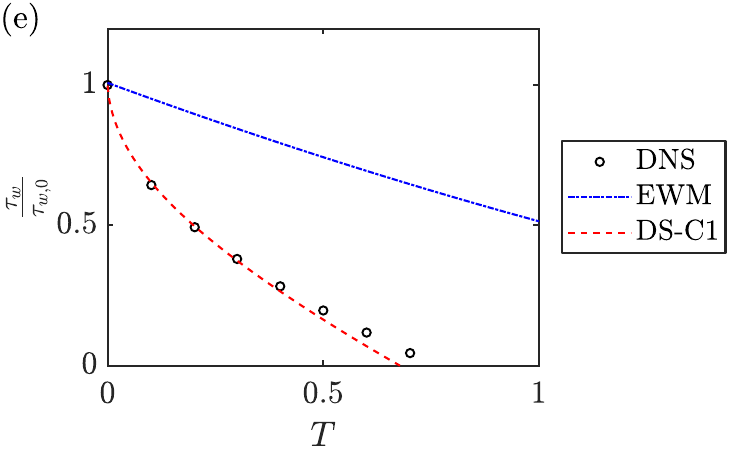}\includegraphics[width=0.32\linewidth]{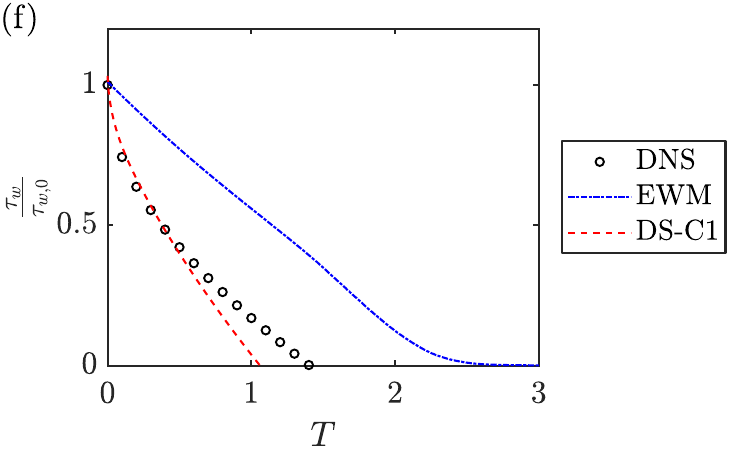}
    \caption{(a-c) Inner-scaled mean velocity profiles following the imposition of an APG: (a) R5APG1; (b) R5APG100; (c) R10APG100. Time $T$ is normalized by $h/U_{c,0}$, where $U_{c,0}$ is the channel centerline velocity at $t=0$. (d-f) Evolution of the wall shear stress: (d) R5APG1; (e) R5APG100; (f) R10APG100. The dashed line corresponds to the results fo the EWM. DNS reference data are shown in color, and predictions from the model in Eq.~\eqref{eq:A22} are labeled `DS-C1'.}
    \label{fig:results_case1}
\end{figure}

We proceed to the {\it a posteriori} test. Figure~\ref{fig:results_case1} presents the results; the computational setup has already been summarized in Table~\ref{tab:setup}. Panels (a–c) show inner-scaled mean velocity profiles at various time instants following the application of the APG, under (a) mild and (b,c) strong APG conditions at two Reynolds numbers. As expected, the EWM performs adequately under mild APG but deteriorates under strong APG, regardless of Reynolds number. In contrast, the model in Eq.~\eqref{eq:A22} accurately tracks the evolution of the mean flow.
Panels (d–f) show the evolution of wall shear stress over time for the three cases shown in panels (a–c). Again, while the EWM is reasonably accurate under mild APG (panel d), it fails to capture the rapid decline in wall shear stress observed in the strong APG cases (panels e–f). By comparison, the DeepSeek model in Eq.~\eqref{eq:A22} consistently produces more accurate wall shear stress predictions across all three cases.
Table~\ref{tab:Quantified_Errors_C1} further quantifies the time to incipient separation for each case. It is evident that the new model substantially reduces the prediction error compared to the baseline EWM.

\begin{table}
    \centering
    \caption{Time to incipient separation (normalized by $h/U_{c,0}$). Relative errors compared to DNS are also listed.}
    \label{tab:Quantified_Errors_C1}
    \begin{tabular}{cccc}
    \hline
         \textbf{Case} & \textbf{DNS} & \textbf{EWM} & \textbf{DS-C1} \\
    \hline
        R5APG1 & $228.4$ & $264.5$ (+15.8\%) & $210.6$ (-7.79\%) \\
        R5APG10 & $22.90$ & $30.71$ (+34.1\%) & $19.30$ (-15.7\%) \\
        R5APG100 & $0.7644$ & $3.086$ (+303.7\%) & $0.6790$ (-11.2\%) \\
        R10APG10 & $29.15$ & $38.14$ (+30.8\%) & $26.06$ (-10.6\%) \\
        R10APG100 & $1.393$ & $3.743$ (+168.7\%) & $1.148$ (-17.6\%) \\
    \hline
    \end{tabular}    
\end{table}

\subsection{System Rotation}
\label{sec2_2:Case2}

System rotation profoundly modifies the dynamics of wall-bounded turbulent flows and arises in a wide range of applications, including turbomachinery, atmospheric flows, and rotating devices. A special case that has received much attention is when the flow is subjected to spanwise system rotation, where the Coriolis force induces a wall-normal pressure gradient that leads to the formation of a ``pressure side'' and a ``suction side'' with turbulence suppressed and enhanced, respectively \citep{johnston1972effects, xia2016direct}. This redistribution of turbulence intensity results in substantial deviations from the classical law-of-the-wall behavior: on the pressure side, the mean velocity exhibits a near-linear profile instead of the canonical logarithmic scaling \citep{yang2020mean,brethouwer2017statistics}.
Consequently, the EWM no longer applies.

\begin{figure}
    \centering
    \includegraphics[width=1.0\linewidth]{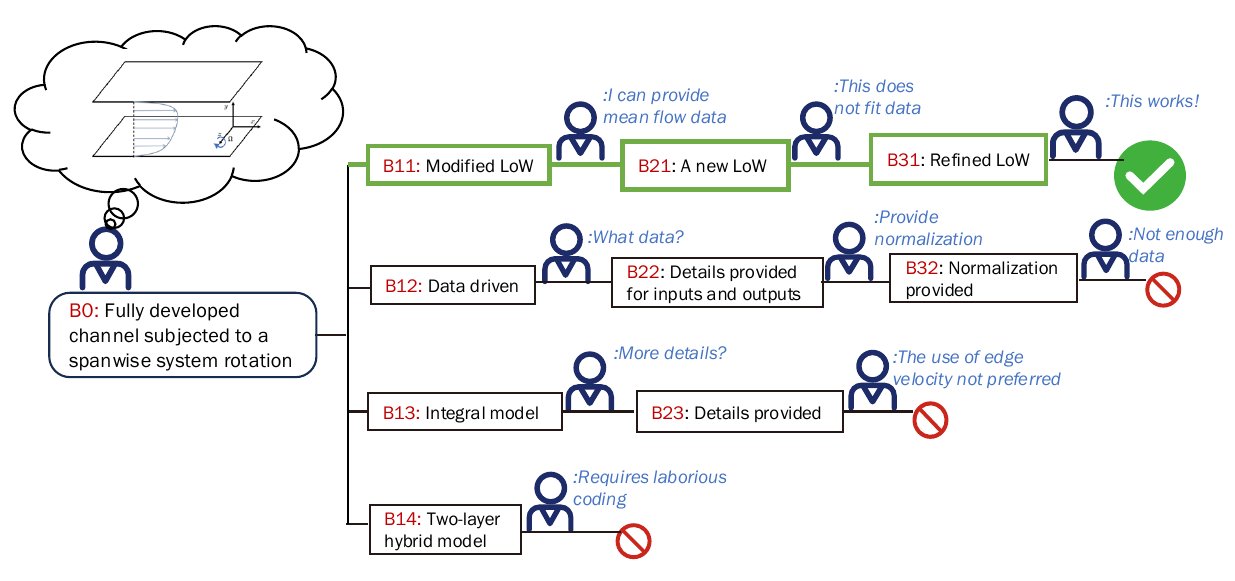}
    \caption{Schematic overview of the interaction between the LLM and the user for the spanwise rotation modeling problem.}
    \label{fig:rotsketch}
\end{figure}

In this section, we engage the LLM to address this challenge. The modeling task was presented to DeepSeek-R1 using the following prompt:
\begin{quote}
\textit{
I am a fluid engineer using wall-modeled large-eddy simulation.  
Turbulent flow in a rotating system has long been a challenging problem. 
I am currently considering a simplified case: a fully developed plane channel flow between two infinitely large plates, where $x$, $y$, and $z$ denote the streamwise, wall-normal, and spanwise directions, respectively. A uniform volume force is applied along the $x$ direction, and the system rotates about the $z$ axis at an angular velocity $\Omega$.  
Are there any existing wall models that can handle this type of flow?}
\end{quote}
An overview of the interaction between the LLM and the user is illustrated in Figure~\ref{fig:rotsketch}.

Before attempting a solution, DeepSeek-R1 undertakes an extended chain of reasoning. It first analyzes the mean flow by invoking and simplifying the RANS equations, concluding that spanwise mean velocities or streamwise vortices may emerge due to Coriolis-Reynolds stress interactions. It correctly concludes that spanwise system rotation modifies turbulence anisotropically: turbulence is suppressed on the stable (suction) side and enhanced on the unstable (pressure) side. Additionally, the altered Reynolds stresses redistribute momentum, fundamentally modifying the mean flow structure.
DeepSeek-R1 then analyzes the limitations of existing wall models. It reasons that the Coriolis force disrupts the equilibrium state assumed in traditional EWMs. Rotation selectively suppresses or enhances turbulent fluctuations, and conventional models are unable to capture such anisotropic effects.
Due to this extended reasoning chain, including the discussion on modeling strategies, the conversation uses approximately 21,000 tokens.

Following this analysis, DeepSeek-R1 proposes several possible solutions. The overarching principles are similar to those observed in the APG problem. A data-driven approach (labeled B12 in Figure~\ref{fig:rotsketch}) is proposed but ultimately discarded due to the lack of high-fidelity training data. An integral formulation (B13) is also suggested but rejected because it requires information at the boundary-layer edge.
What we find particularly interesting in this interaction is that the LLM identifies transport-equation-based models with rotation corrections (B14), which we were not previously aware of and which could potentially form the basis of an effective two-layer model but is not selected due to the significant amount of coding required.
The selected solution (B11) involves modifying the law of the wall. Initially, DeepSeek-R1 recommends altering the logarithmic law. Upon examining the mean flow data, DeepSeek-R1 proposes a modified law of the wall:
{\small\begin{equation}
U^+ = 2 Ro^+ y^+ + ({16.5 + 60.6 Ro^+})/({1 + 46 Ro^+}),
\end{equation}}where $Ro^+$ is the rotation number based on wall units. Inverting the mean flow scaling in Eq.~\eqref{eq:B11} yields a near-wall model:
{\small\begin{equation}
\tau_w/\rho=\left[\frac{U_{\rm LES}}{2 Ro^+ y^+ + ({16.5 + 60.6 Ro^+})/({1 + 46 Ro^+})}\right]^2.
\label{eq:B11}
\end{equation}}Note that, due to the use of $u_\tau$ for non-dimensionalization, evaluating the right-hand side requires knowledge of the wall shear stress, making Eq.~\eqref{eq:B11} an implicit equation for $\tau_w$. Nonetheless, one can avoid an iterative solution by using the wall shear stress from the previous time step to evaluate the right-hand side, following the practice in \cite{yang2017log}, \cite{yang2015integral}, among others.

\begin{figure}
    \centering
    \includegraphics[width=0.23\linewidth]{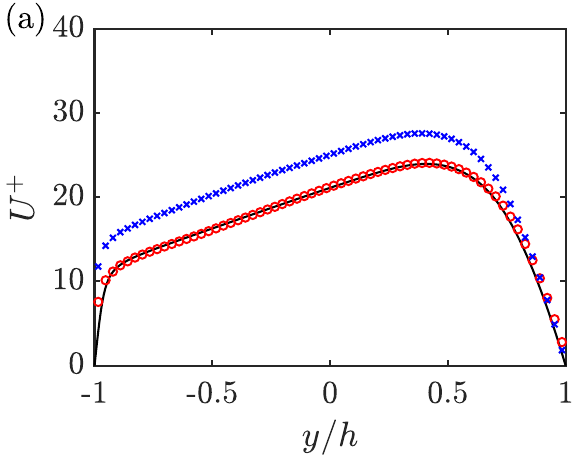}\includegraphics[width=0.23\linewidth]{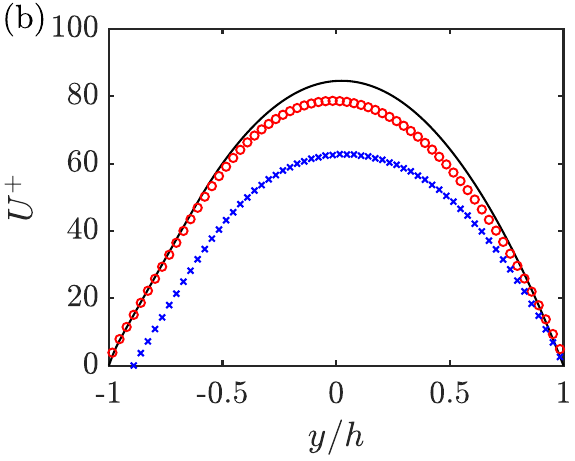}\includegraphics[width=0.23\linewidth]{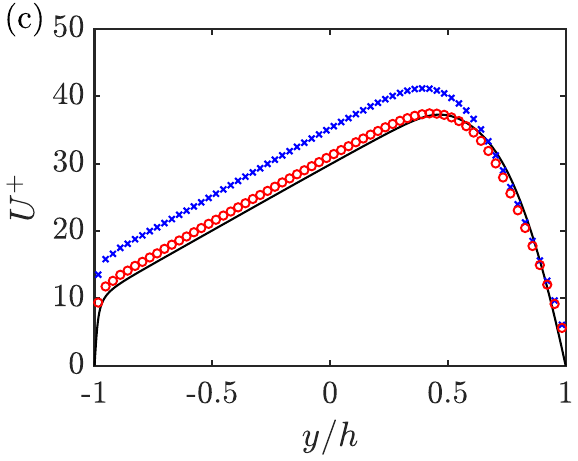}\includegraphics[width=0.29\linewidth]{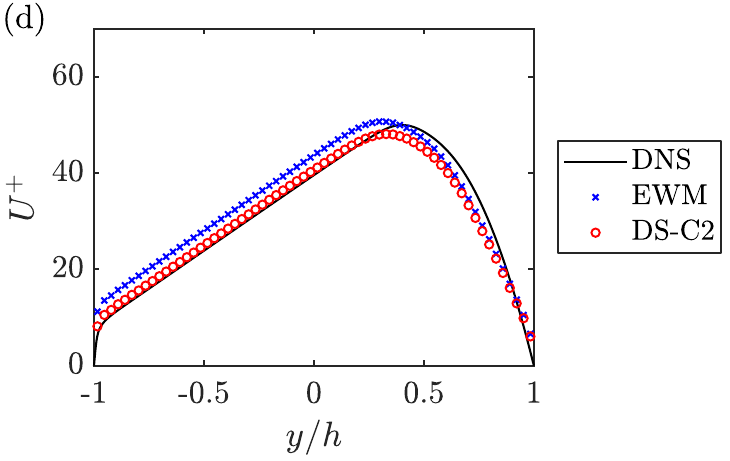}
    \caption{Mean velocity profiles in spanwise rotating channels. (\textit{a}) R2ROT10; (\textit{b}) R2ROT120; (\textit{c}) R4ROT20; (\textit{d}) R4ROT32. Label `DS-C2' corresponds to the model in Eq. \eqref{eq:B11}.}
    \label{fig:results_case3}
\end{figure}

\begin{table}
    \centering
    \caption{Bulk velocity predictions normalized by $u_\tau$. Relative errors compared to DNS are also listed.}
    \label{tab:Quantified_Errors_C2}
    {\small\begin{tabular}{cccc}
    \hline
        \textbf{Case} & \textbf{DNS} & \textbf{EWM} & \textbf{DS-C2} \\
    \hline
        R2ROT10 & $17.62$ & $21.16$ (+20.1\%) & $17.86$ (+1.36\%) \\
        R2ROT22 & $22.25$ & $25.91$ (+16.4\%) & $23.89$ (+7.37\%) \\
        %R2ROT40 & $30.29$ & $31.54$ (+4.13\%) & $34.00$ (+12.2\%) \\
        R2ROT80 & $45.56$ & $39.10$ (-14.2\%) & $47.83$ (+4.98\%) \\
        R2ROT120 & $55.40$ & $39.60$ (-28.5\%) & $52.03$ (-6.08\%) \\
        R4ROT10 & $20.25$ & $24.69$ (+21.9\%) & $21.66$ (+6.96\%) \\
        R4ROT20 & $24.86$ & $28.88$ (+16.2\%) & $25.57$ (+2.86\%) \\
        R4ROT32 & $31.24$ & $33.41$ (+6.95\%) & $30.93$ (-0.99\%) \\
        R4ROT44 & $37.84$ & $37.72$ (-0.32\%) & $37.44$ (-1.06\%) \\
        \hline
    \end{tabular}}
\end{table}

We test the model in Eq.~\eqref{eq:B11} within the WMLES framework for plane channel flows subjected to various levels of spanwise rotation. Figure~\ref{fig:results_case3} presents the mean velocity profiles for four representative cases. For comparison, we include results from the EWM and reference DNS data. The model in Eq.~\eqref{eq:B11} consistently outperforms the baseline EWM and accurately captures the mean flow behavior in spanwise rotating channels.
Additional results are provided in Table~\ref{tab:Quantified_Errors_C2}, listing the predicted bulk velocities normalized by $u_\tau$. Note that a pressure gradient is imposed, the friction Reynolds number is fixed, and the bulk velocity emerges as a prediction of the model. Across all cases, the proposed model significantly improves bulk velocity predictions relative to the EWM over a broad range of rotation numbers.

\subsection{Surface Roughness}
\label{sec2_3:Case3}

Surface roughness significantly impacts boundary-layer flows, altering momentum transfer and introducing substantial drag penalties across a wide range of engineering applications, including ships, aircraft, turbomachinery, and atmospheric boundary layers \citep{barlow2008review,bons2010review,schultz2007effects}. Predicting the influence of roughness has been a longstanding challenge due to the diversity of roughness topographies and their complex interactions with near-wall turbulence \citep{chung2021predicting}. The objective of rough-wall modeling is typically to predict quantities such as the equivalent sand-grain roughness height \( k_s \) or the roughness function \( \Delta U^+ \), and significant advances have been made since the seminal work of Nikuradse \citep{nikuradse1950laws, yang2023search, colebrook1939correspondence, moody1944friction, flack2014roughness}. However, incorporating these insights directly into wall modeling for LES remains limited \citep{bose2018wall}. Here, we engage DeepSeek-R1 to explore new modeling strategies for rough-wall effects in the context of WMLES.
We initiated the discussion with DeepSeek-R1 by posing a general question:
\begin{quote}
\textit{I am a fluid engineer, and I use wall-modeled large-eddy simulation. Modeling surface roughness has always been a difficult problem. What are the main approaches to rough-wall modeling? How can we accommodate surface roughness in wall models?}
\end{quote}

\begin{figure}
    \centering
    \includegraphics[width=1.0\linewidth]{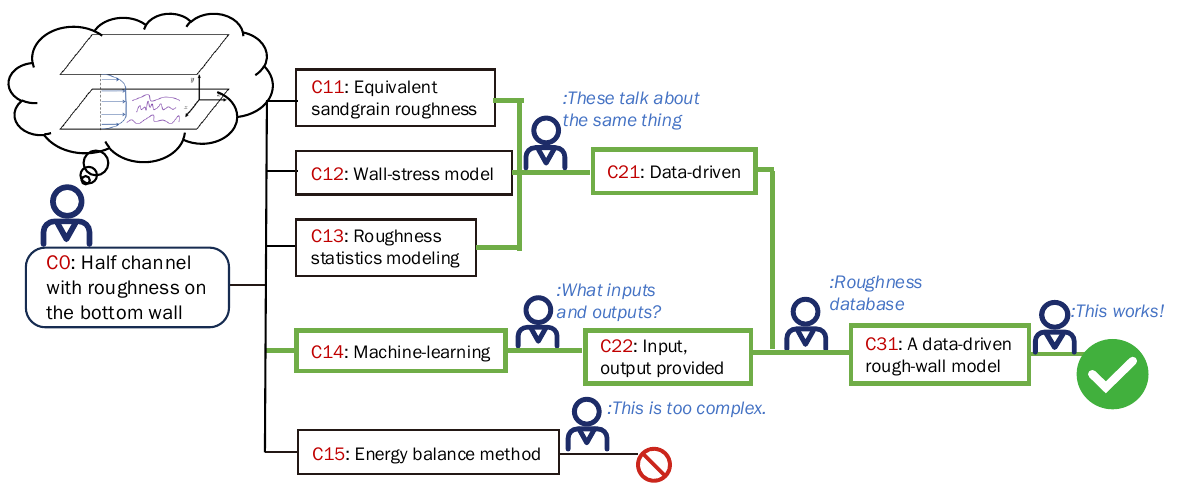}
    \caption{Schematic overview of the interaction between the LLM and the user for the roughness modeling problem.}
    \label{fig:roughsketch}
\end{figure}

DeepSeek-R1 has an awareness of the extensive literature on rough-wall turbulence and highlights several key points.  
First, it notes that the effect of surface roughness is often parameterized by a roughness function, which manifests as a downward shift of the log-law velocity profile relative to smooth-wall flows.  
Second, it recognizes that explicit geometric resolution of roughness in CFD simulations can accurately capture roughness effects, but doing that can be computationally costly.  
Third, it identifies machine-learning approaches that relate near-wall flow information to wall shear stress.  
%Finally, it pointed out that roughness enhances both turbulent kinetic energy and dissipation in the near-wall region, suggesting that incorporating these effects into subgrid-scale modeling could be beneficial.

The full conversation involves approximately 24,000 tokens, and the portion relevant to modeling is summarized schematically in Figure~\ref{fig:roughsketch}.  
In this dialogue, DeepSeek-R1 identifies three key variables for rough-wall modeling: the equivalent sandgrain roughness \( k_s \) (labeled C11), wall shear stress \( \tau_w \) (C12), and parameterizations of the roughness geometry (C13). 
When prompted on how to determine these quantities, the LLM recommends a data-driven approach (C14).
The training data include drag measurements, roughness geometric statistics, and equivalent sandgrain roughness, which are available from public roughness datasets \citep{yang2023search}.  
Upon confirming the feasibility of this data-driven approach, DeepSeek-R1 provides detailed guidance on constructing a neural network, selecting input and output features, and performing the training.
The resulting wall model takes the form:
{\small\begin{equation}
\tau_w/\rho = \left[\frac{U_{\parallel}}{\ln(h_{\rm wm}/k_s)/\kappa + A}\right]^2, \quad k_s = \mathrm{ANN}(\text{Parameterization of Roughness Geometry}),
\label{eq:C31}
\end{equation}}where the parameterization of roughness geometry involves single-point statistics such as the peak-to-trough height, second, third, and fourth-order moments of the roughness height, as well as combinations of these variables. 
We note that comparing the model in Eq.~\eqref{eq:C31} to the EWM is inappropriate, as the EWM assumes a smooth flat plate. 
As a baseline for comparison, we adopt the rough-wall model proposed by \citet{forooghi2017toward}, which is otherwise identical to Eq.~\eqref{eq:C31} except that \( k_s \) is given by:
{\small\begin{equation}
k_s/k_{\rm rms} = 3.41(1 + {\rm SK})^{0.61},
\end{equation}}where \( k_{\rm rms} \) is the root-mean-square of the roughness height, and SK denotes the skewness.

\begin{figure}
    \centering
    \includegraphics[width=0.23\linewidth]{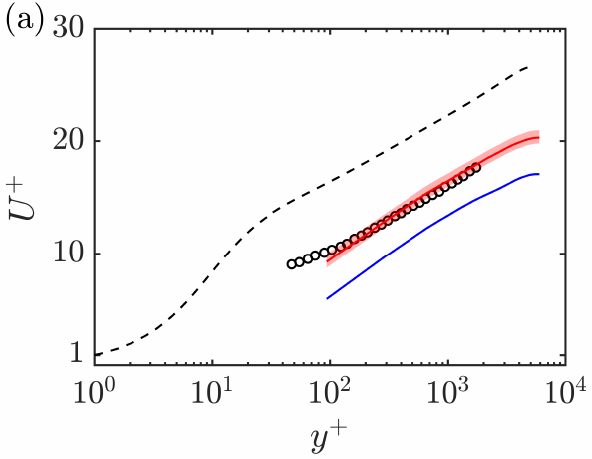}\includegraphics[width=0.23\linewidth]{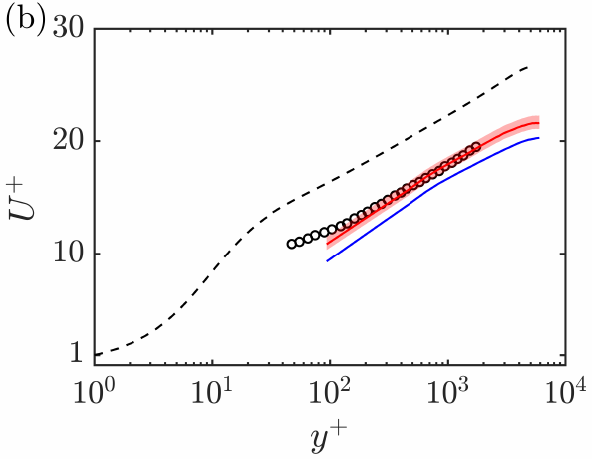}\includegraphics[width=0.23\linewidth]{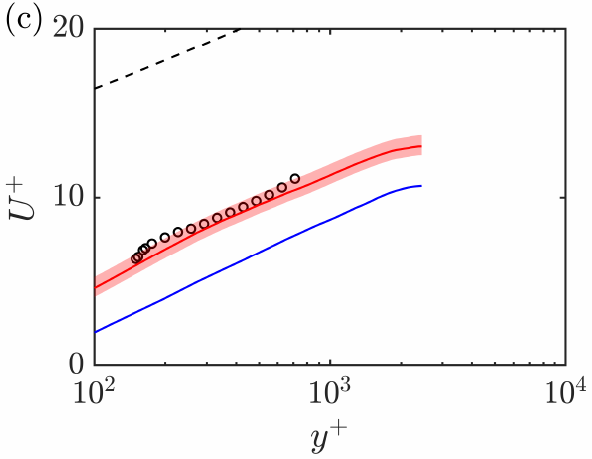}\includegraphics[width=0.29\linewidth]{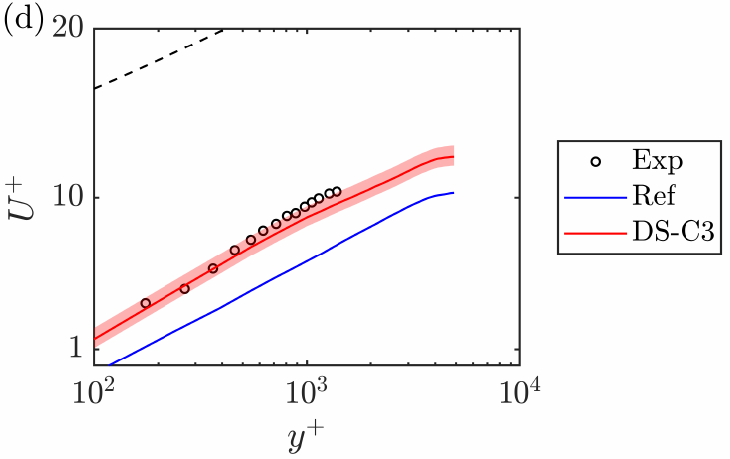}
    \caption{Predicted mean velocity profiles for rough-wall flows: (\textit{a}) R6RW1; (\textit{b}) R6RW3; (\textit{c}) R25RW7; (\textit{d}) R50RW8. 
    %Experimental data are shown in the log region only ($y^+>30$, $y<0.3\delta$). 
    `Ref' corresponds to the roughness model in \cite{forooghi2017toward}. `DS-C3' corresponds to the model in Eq. \eqref{eq:C31}. The dashed black line indicates the LoW. The shaded regions represents the uncertainty in the training data.}
    \label{fig:results_case2}
\end{figure}

We evaluate the performance of the proposed model within the WMLES framework.  
Here, the surface roughness is not explicitly resolved; instead, its effects on the flow are entirely modeled. 
Figure~\ref{fig:results_case2} presents comparisons of mean velocity profiles for several rough-wall cases.  
Results are compared against experimental reference data from \cite{flack2016skin,medjnoun2021turbulent, womack2022turbulent,flack2023hydraulic}, as well as against the baseline model.
The roughness morphologies considered range from random Gaussian surfaces to regular truncated cones.
We observe that the model proposed by DeepSeek-R1 consistently yields more accurate mean velocity predictions, particularly in the logarithmic layer.
More quantitative comparisons are provided in Table~\ref{tab:Quantified_Errors_C3}, which lists the predicted roughness function \( \Delta U^+ \) for all rough surfaces considered. 
The corresponding roughness morphologies are already detailed in Table~\ref{tab:setup}. 
Across all cases, the present model outperforms the reference model of \citet{forooghi2017toward}, highlighting the effectiveness of the LLM-guided modeling approach.

\begin{table}
    \centering
    \caption{Predicted roughness function \( \Delta U^+ \) and the roughness functions measured from experiments. Relative errors compared to experimental measurements are also listed.}
    \label{tab:Quantified_Errors_C3}
    {\small\begin{tabular}{cccc}
        \hline
        \textbf{Case} & \textbf{Exp} & \textbf{Ref} & \textbf{DS-C3} \\
        \hline
        R60RW1 & $6.26$ & $8.80$ (+40.6\%) & $5.75$ (-8.11\%) \\
        R60RW2 & $6.67$ & $9.42$ (+41.2\%) & $6.93$ (+3.86\%) \\
        R60RW3 & $4.43$ & $5.59$ (+26.1\%) & $4.33$ (-2.18\%) \\
        R60RW4 & $4.69$ & $5.87$ (+25.1\%) & $4.10$ (-12.7\%) \\
        R60RW5 & $3.89$ & $4.94$ (+26.9\%) & $3.93$ (+1.06\%) \\
        R60RW6 & $3.30$ & $4.23$ (+28.3\%) & $3.08$ (-6.64\%) \\
        R25RW7 & $10.6$ & $13.3$ (+25.3\%) & $10.66$ (+0.14\%) \\
        R50RW8 & $12.7$ & $15.5$ (+21.4\%) & $13.24$ (+3.96\%) \\
        \hline
    \end{tabular}}
\end{table}

\begin{comment}
\begin{figure}
    \centering
    \includegraphics[width=0.6\linewidth]{statistic.pdf}
    \caption{The proportion of each part throughout the entire dialogue. The percentages of 'Question' and 'instruction' respectively refer to the number of words (in Chinese) produced by humans in problem description and evaluation. 'Rationale' and 'Answer', correspond to the number of words (in Chinese) generated by LLMs in reasoning and answering, respectively.}
    \label{fig:statistic}
\end{figure}
\end{comment}
\begin{comment}
We report the proportion of human-input and AI-output during the dialogue in fig. \ref{fig:statistic}.
DeepSeek-R1 autonomously provided physical reasoning, logical model structures, and outputted model strategies with minor human' evaluation, resulting in models that are transparent, interpretable, and grounded in physical intuition.
\end{comment}

\section{Concluding remarks}
\label{sec:Conclusion}

In this study, we explored the utility of LLMs for turbulence modeling by engaging DeepSeek-R1 in a closed-loop, iterative framework. Within this framework, DeepSeek-R1 engages a human engineers, proposes and refines wall models to address three challenges in near-wall turbulence modeling: APGs, system rotation, and surface roughness. Our results demonstrate that the LLM-driven models not only rival, but in many cases outperform, baseline wall models.

A key distinction between the present work and existing data-driven turbulence modeling efforts lies in the role played by the AI. Conventional data-driven turbulence models are in fact human-driven: researchers design network architectures, curate training data, and employ machine learning to optimize weights and biases. The resulting models often function as black boxes, offering limited interpretability or physical reasoning. In contrast, the models developed in this study were truly AI-driven. 
DeepSeek-R1 autonomously provided physical reasoning, logical model structures, and complete modeling strategies, resulting in models that are transparent, interpretable, and grounded in physical arguments.

Equally important is the paradigm shift in how the AI is treated during the process. In most prior applications, AI tools are treated as subordinate assistants, tasked with executing narrowly defined objectives with minimal feedback loops from human. Here, we treat the LLM as an equal partner, engaging in multiple rounds of iterative dialogue analogous to collaboration between human researchers. Recent studies in human-AI co-creation suggest that enabling reciprocal communication significantly enhances collaboration quality and creative outcomes \citep{rezwana2022identifying}. Our experience supports this view: through sustained interaction, DeepSeek-R1 is able to correct its own errors, refine incomplete formulations, and meaningfully contribute to model development without requiring explicit correction at each step.

The results also highlight a limitation of conventional benchmark evaluations of LLMs, which typically assess models based on a single-response correctness \citep{prohl2024benchmarking,jiang2025deepseek}. As shown here, when allowed to participate in multi-turn, collaborative interactions, LLMs demonstrate capabilities for adaptive reasoning and creative problem-solving that more closely resemble scientific inquiry than simple information retrieval.

Despite these promising results, several limitations of the present investigation must be acknowledged. First, the field of LLM development is advancing rapidly. Since the release of DeepSeek-R1, several newer models have demonstrated even stronger reasoning capabilities, making the present work an initial but already dated baseline. Second, while DeepSeek-R1 proposed models with strong performance and clear logic, it did not invent entirely novel modeling paradigms. The strategies it identified—including data-driven approaches, integral methods, and modified laws of the wall—have precedents in the turbulence modeling literature. Nevertheless, it is important to recognize that individual researchers are typically familiar with only a subset of these methods. The ability of the LLM to synthesize diverse approaches and reason across multiple frameworks represents a significant augmentation of human expertise.

In all, our experiences point to a broader opportunity in applied fluid mechanics: LLMs can serve as powerful collaborators in engineering model developments. They offer a platform for brainstorming ideas, synthesizing knowledge beyond the reach of any individual researcher, and refining concepts through interactive reasoning. 

\paragraph{Funding Statement}
Zhongxin Yang and Yipeng Shi acknowledge NSFC11988102 for financial support.

\paragraph{Declaration of Interests}
The authors declare no conflict of interest.

%\paragraph{Author Contributions}
%The two authors contribute equally to the work.

\paragraph{Data Availability Statement}
Raw data are available from the author upon reasonable request.

\bibliography{BBB_sn-bibliography}% common bib file

%% BioMed_Central_Bib_Style_v1.01

\begin{thebibliography}{91}
% BibTex style file: bmc-mathphys.bst (version 2.1), 2014-07-24
\ifx \bisbn   \undefined \def \bisbn  #1{ISBN #1}\fi
\ifx \binits  \undefined \def \binits#1{#1}\fi
\ifx \bauthor  \undefined \def \bauthor#1{#1}\fi
\ifx \batitle  \undefined \def \batitle#1{#1}\fi
\ifx \bjtitle  \undefined \def \bjtitle#1{#1}\fi
\ifx \bvolume  \undefined \def \bvolume#1{\textbf{#1}}\fi
\ifx \byear  \undefined \def \byear#1{#1}\fi
\ifx \bissue  \undefined \def \bissue#1{#1}\fi
\ifx \bfpage  \undefined \def \bfpage#1{#1}\fi
\ifx \blpage  \undefined \def \blpage #1{#1}\fi
\ifx \burl  \undefined \def \burl#1{\textsf{#1}}\fi
\ifx \doiurl  \undefined \def \doiurl#1{\url{https://doi.org/#1}}\fi
\ifx \betal  \undefined \def \betal{\textit{et al.}}\fi
\ifx \binstitute  \undefined \def \binstitute#1{#1}\fi
\ifx \binstitutionaled  \undefined \def \binstitutionaled#1{#1}\fi
\ifx \bctitle  \undefined \def \bctitle#1{#1}\fi
\ifx \beditor  \undefined \def \beditor#1{#1}\fi
\ifx \bpublisher  \undefined \def \bpublisher#1{#1}\fi
\ifx \bbtitle  \undefined \def \bbtitle#1{#1}\fi
\ifx \bedition  \undefined \def \bedition#1{#1}\fi
\ifx \bseriesno  \undefined \def \bseriesno#1{#1}\fi
\ifx \blocation  \undefined \def \blocation#1{#1}\fi
\ifx \bsertitle  \undefined \def \bsertitle#1{#1}\fi
\ifx \bsnm \undefined \def \bsnm#1{#1}\fi
\ifx \bsuffix \undefined \def \bsuffix#1{#1}\fi
\ifx \bparticle \undefined \def \bparticle#1{#1}\fi
\ifx \barticle \undefined \def \barticle#1{#1}\fi
\bibcommenthead
\ifx \bconfdate \undefined \def \bconfdate #1{#1}\fi
\ifx \botherref \undefined \def \botherref #1{#1}\fi
\ifx \url \undefined \def \url#1{\textsf{#1}}\fi
\ifx \bchapter \undefined \def \bchapter#1{#1}\fi
\ifx \bbook \undefined \def \bbook#1{#1}\fi
\ifx \bcomment \undefined \def \bcomment#1{#1}\fi
\ifx \oauthor \undefined \def \oauthor#1{#1}\fi
\ifx \citeauthoryear \undefined \def \citeauthoryear#1{#1}\fi
\ifx \endbibitem  \undefined \def \endbibitem {}\fi
\ifx \bconflocation  \undefined \def \bconflocation#1{#1}\fi
\ifx \arxivurl  \undefined \def \arxivurl#1{\textsf{#1}}\fi
\csname PreBibitemsHook\endcsname

%%% 1
\bibitem[\protect\citeauthoryear{Silver et~al.}{2017}]{silver2017mastering}
\begin{barticle}
\bauthor{\bsnm{Silver}, \binits{D.}},
\bauthor{\bsnm{Schrittwieser}, \binits{J.}},
\bauthor{\bsnm{Simonyan}, \binits{K.}},
\bauthor{\bsnm{Antonoglou}, \binits{I.}},
\bauthor{\bsnm{Huang}, \binits{A.}},
\bauthor{\bsnm{Guez}, \binits{A.}},
\bauthor{\bsnm{Hubert}, \binits{T.}},
\bauthor{\bsnm{Baker}, \binits{L.}},
\bauthor{\bsnm{Lai}, \binits{M.}},
\bauthor{\bsnm{Bolton}, \binits{A.}}, \betal:
\batitle{Mastering the game of go without human knowledge}.
\bjtitle{Nature}
\bvolume{550}(\bissue{7676}),
\bfpage{354}--\blpage{359}
(\byear{2017})
\end{barticle}
\endbibitem

%%% 2
\bibitem[\protect\citeauthoryear{Jumper et~al.}{2021}]{jumper2021highly}
\begin{barticle}
\bauthor{\bsnm{Jumper}, \binits{J.}},
\bauthor{\bsnm{Evans}, \binits{R.}},
\bauthor{\bsnm{Pritzel}, \binits{A.}},
\bauthor{\bsnm{Green}, \binits{T.}},
\bauthor{\bsnm{Figurnov}, \binits{M.}},
\bauthor{\bsnm{Ronneberger}, \binits{O.}},
\bauthor{\bsnm{Tunyasuvunakool}, \binits{K.}},
\bauthor{\bsnm{Bates}, \binits{R.}},
\bauthor{\bsnm{{\v{Z}}{\'\i}dek}, \binits{A.}},
\bauthor{\bsnm{Potapenko}, \binits{A.}}, \betal:
\batitle{Highly accurate protein structure prediction with alphafold}.
\bjtitle{Nature}
\bvolume{596}(\bissue{7873}),
\bfpage{583}--\blpage{589}
(\byear{2021})
\end{barticle}
\endbibitem

%%% 3
\bibitem[\protect\citeauthoryear{Yurtsever et~al.}{2020}]{yurtsever2020survey}
\begin{barticle}
\bauthor{\bsnm{Yurtsever}, \binits{E.}},
\bauthor{\bsnm{Lambert}, \binits{J.}},
\bauthor{\bsnm{Carballo}, \binits{A.}},
\bauthor{\bsnm{Takeda}, \binits{K.}}:
\batitle{A survey of autonomous driving: Common practices and emerging technologies}.
\bjtitle{IEEE access}
\bvolume{8},
\bfpage{58443}--\blpage{58469}
(\byear{2020})
\end{barticle}
\endbibitem

%%% 4
\bibitem[\protect\citeauthoryear{Buehler}{2024}]{buehler2024mechgpt}
\begin{barticle}
\bauthor{\bsnm{Buehler}, \binits{M.J.}}:
\batitle{{MechGPT}, a language-based strategy for mechanics and materials modeling that connects knowledge across scales, disciplines, and modalities}.
\bjtitle{Applied Mechanics Reviews}
\bvolume{76}(\bissue{2}),
\bfpage{021001}
(\byear{2024})
\end{barticle}
\endbibitem

%%% 5
\bibitem[\protect\citeauthoryear{Ni and Buehler}{2024}]{ni2024mechagents}
\begin{barticle}
\bauthor{\bsnm{Ni}, \binits{B.}},
\bauthor{\bsnm{Buehler}, \binits{M.J.}}:
\batitle{{MechAgents}: Large language model multi-agent collaborations can solve mechanics problems, generate new data, and integrate knowledge}.
\bjtitle{Extreme Mechanics Letters}
\bvolume{67},
\bfpage{102131}
(\byear{2024})
\end{barticle}
\endbibitem

%%% 6
\bibitem[\protect\citeauthoryear{Pandey et~al.}{2025}]{pandey2025openfoamgpt}
\begin{botherref}
\oauthor{\bsnm{Pandey}, \binits{S.}},
\oauthor{\bsnm{Xu}, \binits{R.}},
\oauthor{\bsnm{Wang}, \binits{W.}},
\oauthor{\bsnm{Chu}, \binits{X.}}:
{OpenFOAMGPT}: A retrieval-augmented large language model ({LLM}) agent for {OpenFOAM}-based computational fluid dynamics.
Phys. Fluids
\textbf{37}(3)
(2025)
\end{botherref}
\endbibitem

%%% 7
\bibitem[\protect\citeauthoryear{Dong et~al.}{2025}]{dong2025fine}
\begin{botherref}
\oauthor{\bsnm{Dong}, \binits{Z.}},
\oauthor{\bsnm{Lu}, \binits{Z.}},
\oauthor{\bsnm{Yang}, \binits{Y.}}:
Fine-tuning a large language model for automating computational fluid dynamics simulations.
Theor. Appl. Mech. Lett.,
100594
(2025)
\end{botherref}
\endbibitem

%%% 8
\bibitem[\protect\citeauthoryear{Chang et~al.}{2024}]{chang2024survey}
\begin{barticle}
\bauthor{\bsnm{Chang}, \binits{Y.}},
\bauthor{\bsnm{Wang}, \binits{X.}},
\bauthor{\bsnm{Wang}, \binits{J.}},
\bauthor{\bsnm{Wu}, \binits{Y.}},
\bauthor{\bsnm{Yang}, \binits{L.}},
\bauthor{\bsnm{Zhu}, \binits{K.}},
\bauthor{\bsnm{Chen}, \binits{H.}},
\bauthor{\bsnm{Yi}, \binits{X.}},
\bauthor{\bsnm{Wang}, \binits{C.}},
\bauthor{\bsnm{Wang}, \binits{Y.}}, \betal:
\batitle{A survey on evaluation of large language models}.
\bjtitle{Trans. Intell. Syst. Technol.}
\bvolume{15}(\bissue{3}),
\bfpage{1}--\blpage{45}
(\byear{2024})
\end{barticle}
\endbibitem

%%% 9
\bibitem[\protect\citeauthoryear{Meneveau and Katz}{2000}]{meneveau2000scale}
\begin{barticle}
\bauthor{\bsnm{Meneveau}, \binits{C.}},
\bauthor{\bsnm{Katz}, \binits{J.}}:
\batitle{Scale-invariance and turbulence models for large-eddy simulation}.
\bjtitle{Annu. Rev. Fluid. Mech.}
\bvolume{32}(\bissue{1}),
\bfpage{1}--\blpage{32}
(\byear{2000})
\end{barticle}
\endbibitem

%%% 10
\bibitem[\protect\citeauthoryear{Piomelli and Balaras}{2002}]{piomelli2002wall}
\begin{barticle}
\bauthor{\bsnm{Piomelli}, \binits{U.}},
\bauthor{\bsnm{Balaras}, \binits{E.}}:
\batitle{Wall-layer models for large-eddy simulations}.
\bjtitle{Annu. Rev. Fluid. Mech.}
\bvolume{34}(\bissue{1}),
\bfpage{349}--\blpage{374}
(\byear{2002})
\end{barticle}
\endbibitem

%%% 11
\bibitem[\protect\citeauthoryear{Durbin}{2018}]{durbin2018some}
\begin{barticle}
\bauthor{\bsnm{Durbin}, \binits{P.A.}}:
\batitle{Some recent developments in turbulence closure modeling}.
\bjtitle{Annu. Rev. Fluid. Mech.}
\bvolume{50}(\bissue{1}),
\bfpage{77}--\blpage{103}
(\byear{2018})
\end{barticle}
\endbibitem

%%% 12
\bibitem[\protect\citeauthoryear{Alizadeh}{2022}]{alizadeh2022advances}
\begin{barticle}
\bauthor{\bsnm{Alizadeh}, \binits{O.}}:
\batitle{Advances and challenges in climate modeling}.
\bjtitle{Climatic Change}
\bvolume{170}(\bissue{1}),
\bfpage{18}
(\byear{2022})
\end{barticle}
\endbibitem

%%% 13
\bibitem[\protect\citeauthoryear{Mani and Dorgan}{2023}]{mani2023perspective}
\begin{barticle}
\bauthor{\bsnm{Mani}, \binits{M.}},
\bauthor{\bsnm{Dorgan}, \binits{A.J.}}:
\batitle{A perspective on the state of aerospace computational fluid dynamics technology}.
\bjtitle{Annu. Rev. Fluid. Mech.}
\bvolume{55}(\bissue{1}),
\bfpage{431}--\blpage{457}
(\byear{2023})
\end{barticle}
\endbibitem

%%% 14
\bibitem[\protect\citeauthoryear{Stevens and Meneveau}{2017}]{stevens2017flow}
\begin{barticle}
\bauthor{\bsnm{Stevens}, \binits{R.J.}},
\bauthor{\bsnm{Meneveau}, \binits{C.}}:
\batitle{Flow structure and turbulence in wind farms}.
\bjtitle{Annu. Rev. Fluid. Mech.}
\bvolume{49}(\bissue{1}),
\bfpage{311}--\blpage{339}
(\byear{2017})
\end{barticle}
\endbibitem

%%% 15
\bibitem[\protect\citeauthoryear{Yang and Griffin}{2021}]{yang2021grid}
\begin{botherref}
\oauthor{\bsnm{Yang}, \binits{X.I.A.}},
\oauthor{\bsnm{Griffin}, \binits{K.P.}}:
Grid-point and time-step requirements for direct numerical simulation and large-eddy simulation.
Phys. Fluids
\textbf{33}(1)
(2021)
\end{botherref}
\endbibitem

%%% 16
\bibitem[\protect\citeauthoryear{Choi and Moin}{2012}]{choi2012grid}
\begin{botherref}
\oauthor{\bsnm{Choi}, \binits{H.}},
\oauthor{\bsnm{Moin}, \binits{P.}}:
Grid-point requirements for large eddy simulation: Chapman’s estimates revisited.
Phys. Fluids
\textbf{24}(1)
(2012)
\end{botherref}
\endbibitem

%%% 17
\bibitem[\protect\citeauthoryear{Goc et~al.}{2021}]{goc2021large}
\begin{barticle}
\bauthor{\bsnm{Goc}, \binits{K.A.}},
\bauthor{\bsnm{Lehmkuhl}, \binits{O.}},
\bauthor{\bsnm{Park}, \binits{G.I.}},
\bauthor{\bsnm{Bose}, \binits{S.T.}},
\bauthor{\bsnm{Moin}, \binits{P.}}:
\batitle{Large eddy simulation of aircraft at affordable cost: a milestone in computational fluid dynamics}.
\bjtitle{Flow}
\bvolume{1},
\bfpage{14}
(\byear{2021})
\end{barticle}
\endbibitem

%%% 18
\bibitem[\protect\citeauthoryear{Goc et~al.}{2024}]{goc2024wind}
\begin{barticle}
\bauthor{\bsnm{Goc}, \binits{K.A.}},
\bauthor{\bsnm{Moin}, \binits{P.}},
\bauthor{\bsnm{Bose}, \binits{S.T.}},
\bauthor{\bsnm{Clark}, \binits{A.M.}}:
\batitle{Wind tunnel and grid resolution effects in large-eddy simulations of the high-lift common research model}.
\bjtitle{J Aircr}
\bvolume{61}(\bissue{1}),
\bfpage{267}--\blpage{279}
(\byear{2024})
\end{barticle}
\endbibitem

%%% 19
\bibitem[\protect\citeauthoryear{Marusic and Monty}{2019}]{marusic2019attached}
\begin{barticle}
\bauthor{\bsnm{Marusic}, \binits{I.}},
\bauthor{\bsnm{Monty}, \binits{J.P.}}:
\batitle{Attached eddy model of wall turbulence}.
\bjtitle{Annu. Rev. Fluid. Mech.}
\bvolume{51}(\bissue{1}),
\bfpage{49}--\blpage{74}
(\byear{2019})
\end{barticle}
\endbibitem

%%% 20
\bibitem[\protect\citeauthoryear{Bose and Park}{2018}]{bose2018wall}
\begin{barticle}
\bauthor{\bsnm{Bose}, \binits{S.T.}},
\bauthor{\bsnm{Park}, \binits{G.I.}}:
\batitle{Wall-modeled large-eddy simulation for complex turbulent flows}.
\bjtitle{Annu. Rev. Fluid. Mech.}
\bvolume{50}(\bissue{1}),
\bfpage{535}--\blpage{561}
(\byear{2018})
\end{barticle}
\endbibitem

%%% 21
\bibitem[\protect\citeauthoryear{Larsson et~al.}{2016}]{larsson2016large}
\begin{barticle}
\bauthor{\bsnm{Larsson}, \binits{J.}},
\bauthor{\bsnm{Kawai}, \binits{S.}},
\bauthor{\bsnm{Bodart}, \binits{J.}},
\bauthor{\bsnm{Bermejo-Moreno}, \binits{I.}}:
\batitle{Large eddy simulation with modeled wall-stress: recent progress and future directions}.
\bjtitle{Mechanical Engineering Reviews}
\bvolume{3}(\bissue{1}),
\bfpage{15}--\blpage{00418}
(\byear{2016})
\end{barticle}
\endbibitem

%%% 22
\bibitem[\protect\citeauthoryear{Schumann}{1975}]{schumann1975subgrid}
\begin{barticle}
\bauthor{\bsnm{Schumann}, \binits{U.}}:
\batitle{Subgrid scale model for finite difference simulations of turbulent flows in plane channels and annuli}.
\bjtitle{J. Comput. Phys.}
\bvolume{18}(\bissue{4}),
\bfpage{376}--\blpage{404}
(\byear{1975})
\end{barticle}
\endbibitem

%%% 23
\bibitem[\protect\citeauthoryear{Kawai and Larsson}{2012}]{kawai2012wall}
\begin{botherref}
\oauthor{\bsnm{Kawai}, \binits{S.}},
\oauthor{\bsnm{Larsson}, \binits{J.}}:
Wall-modeling in large eddy simulation: Length scales, grid resolution, and accuracy.
Phys. Fluids
\textbf{24}(1)
(2012)
\end{botherref}
\endbibitem

%%% 24
\bibitem[\protect\citeauthoryear{Yang et~al.}{2017}]{yang2017log}
\begin{barticle}
\bauthor{\bsnm{Yang}, \binits{X.I.A.}},
\bauthor{\bsnm{Park}, \binits{G.I.}},
\bauthor{\bsnm{Moin}, \binits{P.}}:
\batitle{Log-layer mismatch and modeling of the fluctuating wall stress in wall-modeled large-eddy simulations}.
\bjtitle{Phys. Rev. Fluids}
\bvolume{2}(\bissue{10}),
\bfpage{104601}
(\byear{2017})
\end{barticle}
\endbibitem

%%% 25
\bibitem[\protect\citeauthoryear{Park and Moin}{2014}]{park2014improved}
\begin{botherref}
\oauthor{\bsnm{Park}, \binits{G.I.}},
\oauthor{\bsnm{Moin}, \binits{P.}}:
An improved dynamic non-equilibrium wall-model for large eddy simulation.
Phys. Fluids
\textbf{26}(1)
(2014)
\end{botherref}
\endbibitem

%%% 26
\bibitem[\protect\citeauthoryear{Yang et~al.}{2015}]{yang2015integral}
\begin{botherref}
\oauthor{\bsnm{Yang}, \binits{X.I.A.}},
\oauthor{\bsnm{Sadique}, \binits{J.}},
\oauthor{\bsnm{Mittal}, \binits{R.}},
\oauthor{\bsnm{Meneveau}, \binits{C.}}:
Integral wall model for large eddy simulations of wall-bounded turbulent flows.
Phys. Fluids
\textbf{27}(2)
(2015)
\end{botherref}
\endbibitem

%%% 27
\bibitem[\protect\citeauthoryear{Bose and Moin}{2014}]{bose2014dynamic}
\begin{botherref}
\oauthor{\bsnm{Bose}, \binits{S.T.}},
\oauthor{\bsnm{Moin}, \binits{P.}}:
A dynamic slip boundary condition for wall-modeled large-eddy simulation.
Phys. Fluids
\textbf{26}(1)
(2014)
\end{botherref}
\endbibitem

%%% 28
\bibitem[\protect\citeauthoryear{Bae et~al.}{2019}]{bae2019dynamic}
\begin{barticle}
\bauthor{\bsnm{Bae}, \binits{H.J.}},
\bauthor{\bsnm{Lozano-Dur{\'a}n}, \binits{A.}},
\bauthor{\bsnm{Bose}, \binits{S.T.}},
\bauthor{\bsnm{Moin}, \binits{P.}}:
\batitle{Dynamic slip wall model for large-eddy simulation}.
\bjtitle{J. Fluid Mech.}
\bvolume{859},
\bfpage{400}--\blpage{432}
(\byear{2019})
\end{barticle}
\endbibitem

%%% 29
\bibitem[\protect\citeauthoryear{Fowler et~al.}{2022}]{fowler2022lagrangian}
\begin{barticle}
\bauthor{\bsnm{Fowler}, \binits{M.}},
\bauthor{\bsnm{Zaki}, \binits{T.A.}},
\bauthor{\bsnm{Meneveau}, \binits{C.}}:
\batitle{A lagrangian relaxation towards equilibrium wall model for large eddy simulation}.
\bjtitle{J. Fluid Mech.}
\bvolume{934},
\bfpage{44}
(\byear{2022})
\end{barticle}
\endbibitem

%%% 30
\bibitem[\protect\citeauthoryear{Yang et~al.}{2024}]{yang2024predictive}
\begin{barticle}
\bauthor{\bsnm{Yang}, \binits{X.I.A.}},
\bauthor{\bsnm{Chen}, \binits{P.E.}},
\bauthor{\bsnm{Zhang}, \binits{W.}},
\bauthor{\bsnm{Kunz}, \binits{R.}}:
\batitle{Predictive near-wall modelling for turbulent boundary layers with arbitrary pressure gradients}.
\bjtitle{J. Fluid Mech.}
\bvolume{993},
\bfpage{1}
(\byear{2024})
\end{barticle}
\endbibitem

%%% 31
\bibitem[\protect\citeauthoryear{Duraisamy et~al.}{2019}]{duraisamy2019turbulence}
\begin{barticle}
\bauthor{\bsnm{Duraisamy}, \binits{K.}},
\bauthor{\bsnm{Iaccarino}, \binits{G.}},
\bauthor{\bsnm{Xiao}, \binits{H.}}:
\batitle{Turbulence modeling in the age of data}.
\bjtitle{Annu. Rev. Fluid. Mech.}
\bvolume{51}(\bissue{1}),
\bfpage{357}--\blpage{377}
(\byear{2019})
\end{barticle}
\endbibitem

%%% 32
\bibitem[\protect\citeauthoryear{Pandey et~al.}{2020}]{pandey2020perspective}
\begin{barticle}
\bauthor{\bsnm{Pandey}, \binits{S.}},
\bauthor{\bsnm{Schumacher}, \binits{J.}},
\bauthor{\bsnm{Sreenivasan}, \binits{K.R.}}:
\batitle{A perspective on machine learning in turbulent flows}.
\bjtitle{J. Turbul}
\bvolume{21}(\bissue{9-10}),
\bfpage{567}--\blpage{584}
(\byear{2020})
\end{barticle}
\endbibitem

%%% 33
\bibitem[\protect\citeauthoryear{Shan et~al.}{2023}]{shan2023turbulence}
\begin{barticle}
\bauthor{\bsnm{Shan}, \binits{X.}},
\bauthor{\bsnm{Liu}, \binits{Y.}},
\bauthor{\bsnm{Cao}, \binits{W.}},
\bauthor{\bsnm{Sun}, \binits{X.}},
\bauthor{\bsnm{Zhang}, \binits{W.}}:
\batitle{Turbulence modeling via data assimilation and machine learning for separated flows over airfoils}.
\bjtitle{AIAA J.}
\bvolume{61}(\bissue{9}),
\bfpage{3883}--\blpage{3899}
(\byear{2023})
\end{barticle}
\endbibitem

%%% 34
\bibitem[\protect\citeauthoryear{Bin et~al.}{2022}]{bin2022progressive}
\begin{barticle}
\bauthor{\bsnm{Bin}, \binits{Y.}},
\bauthor{\bsnm{Chen}, \binits{L.}},
\bauthor{\bsnm{Huang}, \binits{G.}},
\bauthor{\bsnm{Yang}, \binits{X.I.A.}}:
\batitle{Progressive, extrapolative machine learning for near-wall turbulence modeling}.
\bjtitle{Phys. Rev. Fluids}
\bvolume{7}(\bissue{8}),
\bfpage{084610}
(\byear{2022})
\end{barticle}
\endbibitem

%%% 35
\bibitem[\protect\citeauthoryear{Bin et~al.}{2023}]{bin2023data}
\begin{barticle}
\bauthor{\bsnm{Bin}, \binits{Y.}},
\bauthor{\bsnm{Huang}, \binits{G.}},
\bauthor{\bsnm{Yang}, \binits{X.I.A.}}:
\batitle{Data-enabled recalibration of the {S}palart-{A}llmaras model}.
\bjtitle{AIAA J.}
\bvolume{61}(\bissue{11}),
\bfpage{4852}--\blpage{4863}
(\byear{2023})
\end{barticle}
\endbibitem

%%% 36
\bibitem[\protect\citeauthoryear{Maulik et~al.}{2019}]{maulik2019sub}
\begin{barticle}
\bauthor{\bsnm{Maulik}, \binits{R.}},
\bauthor{\bsnm{San}, \binits{O.}},
\bauthor{\bsnm{Jacob}, \binits{J.D.}},
\bauthor{\bsnm{Crick}, \binits{C.}}:
\batitle{Sub-grid scale model classification and blending through deep learning}.
\bjtitle{J. Fluid Mech.}
\bvolume{870},
\bfpage{784}--\blpage{812}
(\byear{2019})
\end{barticle}
\endbibitem

%%% 37
\bibitem[\protect\citeauthoryear{Cheng et~al.}{2022}]{cheng2022deep}
\begin{barticle}
\bauthor{\bsnm{Cheng}, \binits{Y.}},
\bauthor{\bsnm{Giometto}, \binits{M.G.}},
\bauthor{\bsnm{Kauffmann}, \binits{P.}},
\bauthor{\bsnm{Lin}, \binits{L.}},
\bauthor{\bsnm{Cao}, \binits{C.}},
\bauthor{\bsnm{Zupnick}, \binits{C.}},
\bauthor{\bsnm{Li}, \binits{H.}},
\bauthor{\bsnm{Li}, \binits{Q.}},
\bauthor{\bsnm{Huang}, \binits{Y.}},
\bauthor{\bsnm{Abernathey}, \binits{R.}}, \betal:
\batitle{Deep learning for subgrid-scale turbulence modeling in large-eddy simulations of the convective atmospheric boundary layer}.
\bjtitle{J. Adv. Model. Earth Syst.}
\bvolume{14}(\bissue{5}),
\bfpage{2021}--\blpage{002847}
(\byear{2022})
\end{barticle}
\endbibitem

%%% 38
\bibitem[\protect\citeauthoryear{Xie et~al.}{2020}]{xie2020modeling}
\begin{barticle}
\bauthor{\bsnm{Xie}, \binits{C.}},
\bauthor{\bsnm{Wang}, \binits{J.}},
\bauthor{\bsnm{E}, \binits{W.}}:
\batitle{Modeling subgrid-scale forces by spatial artificial neural networks in large eddy simulation of turbulence}.
\bjtitle{Phys. Rev. Fluids}
\bvolume{5}(\bissue{5}),
\bfpage{054606}
(\byear{2020})
\end{barticle}
\endbibitem

%%% 39
\bibitem[\protect\citeauthoryear{Ling et~al.}{2016}]{ling2016reynolds}
\begin{barticle}
\bauthor{\bsnm{Ling}, \binits{J.}},
\bauthor{\bsnm{Kurzawski}, \binits{A.}},
\bauthor{\bsnm{Templeton}, \binits{J.}}:
\batitle{{R}eynolds averaged turbulence modelling using deep neural networks with embedded invariance}.
\bjtitle{J. Fluid Mech.}
\bvolume{807},
\bfpage{155}--\blpage{166}
(\byear{2016})
\end{barticle}
\endbibitem

%%% 40
\bibitem[\protect\citeauthoryear{Parish and Duraisamy}{2016}]{parish2016paradigm}
\begin{barticle}
\bauthor{\bsnm{Parish}, \binits{E.J.}},
\bauthor{\bsnm{Duraisamy}, \binits{K.}}:
\batitle{A paradigm for data-driven predictive modeling using field inversion and machine learning}.
\bjtitle{J. Comput. Phys.}
\bvolume{305},
\bfpage{758}--\blpage{774}
(\byear{2016})
\end{barticle}
\endbibitem

%%% 41
\bibitem[\protect\citeauthoryear{Wang et~al.}{2017}]{wang2017physics}
\begin{barticle}
\bauthor{\bsnm{Wang}, \binits{J.-X.}},
\bauthor{\bsnm{Wu}, \binits{J.-L.}},
\bauthor{\bsnm{Xiao}, \binits{H.}}:
\batitle{Physics-informed machine learning approach for reconstructing {R}eynolds stress modeling discrepancies based on {DNS} data}.
\bjtitle{Phys. Rev. Fluids}
\bvolume{2}(\bissue{3}),
\bfpage{034603}
(\byear{2017})
\end{barticle}
\endbibitem

%%% 42
\bibitem[\protect\citeauthoryear{Bin et~al.}{2024a}]{bin2024constrainedsa}
\begin{barticle}
\bauthor{\bsnm{Bin}, \binits{Y.}},
\bauthor{\bsnm{Huang}, \binits{G.}},
\bauthor{\bsnm{Kunz}, \binits{R.}},
\bauthor{\bsnm{Yang}, \binits{X.I.A.}}:
\batitle{Constrained recalibration of {R}eynolds-averaged {N}avier-{S}tokes models}.
\bjtitle{AIAA J.}
\bvolume{62}(\bissue{4}),
\bfpage{1434}--\blpage{1446}
(\byear{2024})
\end{barticle}
\endbibitem

%%% 43
\bibitem[\protect\citeauthoryear{Bin et~al.}{2024b}]{bin2024constrainedsst}
\begin{barticle}
\bauthor{\bsnm{Bin}, \binits{Y.}},
\bauthor{\bsnm{Hu}, \binits{X.}},
\bauthor{\bsnm{Li}, \binits{J.}},
\bauthor{\bsnm{Grauer}, \binits{S.J.}},
\bauthor{\bsnm{Yang}, \binits{X.I.A.}}:
\batitle{Constrained re-calibration of two-equation {R}eynolds-averaged {N}avier-{S}tokes models}.
\bjtitle{Theor. Appl. Mech. Lett.}
\bvolume{14}(\bissue{2}),
\bfpage{100503}
(\byear{2024})
\end{barticle}
\endbibitem

%%% 44
\bibitem[\protect\citeauthoryear{Wu et~al.}{2025}]{wu2025development}
\begin{barticle}
\bauthor{\bsnm{Wu}, \binits{C.}},
\bauthor{\bsnm{Zhang}, \binits{S.}},
\bauthor{\bsnm{Zhang}, \binits{Y.}}:
\batitle{Development of a generalizable data-driven turbulence model: Conditioned field inversion and symbolic regression}.
\bjtitle{AIAA J.}
\bvolume{63}(\bissue{2}),
\bfpage{687}--\blpage{706}
(\byear{2025})
\end{barticle}
\endbibitem

%%% 45
\bibitem[\protect\citeauthoryear{Yang et~al.}{2019}]{yang2019predictive}
\begin{barticle}
\bauthor{\bsnm{Yang}, \binits{X.I.A.}},
\bauthor{\bsnm{Zafar}, \binits{S.}},
\bauthor{\bsnm{Wang}, \binits{J.-X.}},
\bauthor{\bsnm{Xiao}, \binits{H.}}:
\batitle{Predictive large-eddy-simulation wall modeling via physics-informed neural networks}.
\bjtitle{Phys. Rev. Fluids}
\bvolume{4}(\bissue{3}),
\bfpage{034602}
(\byear{2019})
\end{barticle}
\endbibitem

%%% 46
\bibitem[\protect\citeauthoryear{Bae and Koumoutsakos}{2022}]{bae2022scientific}
\begin{barticle}
\bauthor{\bsnm{Bae}, \binits{H.J.}},
\bauthor{\bsnm{Koumoutsakos}, \binits{P.}}:
\batitle{Scientific multi-agent reinforcement learning for wall-models of turbulent flows}.
\bjtitle{Nat. Commun.}
\bvolume{13}(\bissue{1}),
\bfpage{1443}
(\byear{2022})
\end{barticle}
\endbibitem

%%% 47
\bibitem[\protect\citeauthoryear{Vadrot et~al.}{2023}]{vadrot2023log}
\begin{botherref}
\oauthor{\bsnm{Vadrot}, \binits{A.}},
\oauthor{\bsnm{Yang}, \binits{X.I.A.}},
\oauthor{\bsnm{Bae}, \binits{H.J.}},
\oauthor{\bsnm{Abkar}, \binits{M.}}:
Log-law recovery through reinforcement-learning wall model for large eddy simulation.
Phys. Fluids
\textbf{35}(5)
(2023)
\end{botherref}
\endbibitem

%%% 48
\bibitem[\protect\citeauthoryear{Zhou et~al.}{2021}]{zhou2021wall}
\begin{barticle}
\bauthor{\bsnm{Zhou}, \binits{Z.}},
\bauthor{\bsnm{He}, \binits{G.}},
\bauthor{\bsnm{Yang}, \binits{X.}}:
\batitle{Wall model based on neural networks for les of turbulent flows over periodic hills}.
\bjtitle{Phys. Rev. Fluids}
\bvolume{6}(\bissue{5}),
\bfpage{054610}
(\byear{2021})
\end{barticle}
\endbibitem

%%% 49
\bibitem[\protect\citeauthoryear{Ma and Lozano-Dur{\'a}n}{2025}]{ma2025machine}
\begin{barticle}
\bauthor{\bsnm{Ma}, \binits{R.}},
\bauthor{\bsnm{Lozano-Dur{\'a}n}, \binits{A.}}:
\batitle{Machine-learning wall-model large-eddy simulation accounting for isotropic roughness under local equilibrium}.
\bjtitle{J. Fluid Mech.}
\bvolume{1007},
\bfpage{17}
(\byear{2025})
\end{barticle}
\endbibitem

%%% 50
\bibitem[\protect\citeauthoryear{Lozano-Dur{\'a}n and Bae}{2023}]{lozano2023machine}
\begin{barticle}
\bauthor{\bsnm{Lozano-Dur{\'a}n}, \binits{A.}},
\bauthor{\bsnm{Bae}, \binits{H.J.}}:
\batitle{Machine learning building-block-flow wall model for large-eddy simulation}.
\bjtitle{J. Fluid Mech.}
\bvolume{963},
\bfpage{35}
(\byear{2023})
\end{barticle}
\endbibitem

%%% 51
\bibitem[\protect\citeauthoryear{Vadrot et~al.}{2023}]{vadrot2023survey}
\begin{barticle}
\bauthor{\bsnm{Vadrot}, \binits{A.}},
\bauthor{\bsnm{Yang}, \binits{X.I.A.}},
\bauthor{\bsnm{Abkar}, \binits{M.}}:
\batitle{Survey of machine-learning wall models for large-eddy simulation}.
\bjtitle{Phys. Rev. Fluids}
\bvolume{8}(\bissue{6}),
\bfpage{064603}
(\byear{2023})
\end{barticle}
\endbibitem

%%% 52
\bibitem[\protect\citeauthoryear{Pr{\"o}hl et~al.}{2024}]{prohl2024benchmarking}
\begin{botherref}
\oauthor{\bsnm{Pr{\"o}hl}, \binits{T.}},
\oauthor{\bsnm{Putzier}, \binits{E.}},
\oauthor{\bsnm{Zarnekow}, \binits{R.}}:
Benchmarking of {LLM} detection: Comparing two competing approaches.
arXiv preprint arXiv:2406.11670
(2024)
\end{botherref}
\endbibitem

%%% 53
\bibitem[\protect\citeauthoryear{Jiang et~al.}{2025}]{jiang2025deepseek}
\begin{barticle}
\bauthor{\bsnm{Jiang}, \binits{Q.}},
\bauthor{\bsnm{Gao}, \binits{Z.}},
\bauthor{\bsnm{Karniadakis}, \binits{G.E.}}:
\batitle{Deepseek vs. {ChatGPT} vs. {Claude}: A comparative study for scientific computing and scientific machine learning tasks}.
\bjtitle{Theor. Appl. Mech. Lett.}
\bvolume{15}(\bissue{3}),
\bfpage{100583}
(\byear{2025})
\end{barticle}
\endbibitem

%%% 54
\bibitem[\protect\citeauthoryear{Gao et~al.}{2025}]{gao2025comparison}
\begin{botherref}
\oauthor{\bsnm{Gao}, \binits{T.}},
\oauthor{\bsnm{Jin}, \binits{J.}},
\oauthor{\bsnm{Ke}, \binits{Z.T.}},
\oauthor{\bsnm{Moryoussef}, \binits{G.}}:
A comparison of deepseek and other {LLM}s.
arXiv preprint arXiv:2502.03688
(2025)
\end{botherref}
\endbibitem

%%% 55
\bibitem[\protect\citeauthoryear{Yang et~al.}{2020}]{yang2020mean}
\begin{barticle}
\bauthor{\bsnm{Yang}, \binits{X.I.A.}},
\bauthor{\bsnm{Xia}, \binits{Z.-H.}},
\bauthor{\bsnm{Lee}, \binits{J.}},
\bauthor{\bsnm{Lv}, \binits{Y.}},
\bauthor{\bsnm{Yuan}, \binits{J.}}:
\batitle{Mean flow scaling in a spanwise rotating channel}.
\bjtitle{Phys. Rev. Fluids}
\bvolume{5}(\bissue{7}),
\bfpage{074603}
(\byear{2020})
\end{barticle}
\endbibitem

%%% 56
\bibitem[\protect\citeauthoryear{Chen et~al.}{2023}]{chen2023universal}
\begin{barticle}
\bauthor{\bsnm{Chen}, \binits{P.E.}},
\bauthor{\bsnm{Wu}, \binits{W.}},
\bauthor{\bsnm{Griffin}, \binits{K.P.}},
\bauthor{\bsnm{Shi}, \binits{Y.}},
\bauthor{\bsnm{Yang}, \binits{X.I.A.}}:
\batitle{A universal velocity transformation for boundary layers with pressure gradients}.
\bjtitle{J. Fluid Mech.}
\bvolume{970},
\bfpage{3}
(\byear{2023})
\end{barticle}
\endbibitem

%%% 57
\bibitem[\protect\citeauthoryear{Huang and Yang}{2021}]{huang2021bayesian}
\begin{botherref}
\oauthor{\bsnm{Huang}, \binits{X.L.}},
\oauthor{\bsnm{Yang}, \binits{X.I.A.}}:
A {B}ayesian approach to the mean flow in a channel with small but arbitrarily directional system rotation.
Phys. Fluids
\textbf{33}(1)
(2021)
\end{botherref}
\endbibitem

%%% 58
\bibitem[\protect\citeauthoryear{Yang et~al.}{2023}]{yang2023search}
\begin{botherref}
\oauthor{\bsnm{Yang}, \binits{X.I.A.}},
\oauthor{\bsnm{Zhang}, \binits{W.}},
\oauthor{\bsnm{Yuan}, \binits{J.}},
\oauthor{\bsnm{Kunz}, \binits{R.F.}}:
In search of a universal rough wall model.
J. Fluids Eng.
\textbf{145}(10)
(2023)
\end{botherref}
\endbibitem

%%% 59
\bibitem[\protect\citeauthoryear{Nair et~al.}{2024}]{nair2024rough}
\begin{barticle}
\bauthor{\bsnm{Nair}, \binits{S.S.}},
\bauthor{\bsnm{Wadhai}, \binits{V.A.}},
\bauthor{\bsnm{Kunz}, \binits{R.F.}},
\bauthor{\bsnm{Yang}, \binits{X.I.A.}}:
\batitle{Rough surfaces in underexplored surface morphology space and their implications on roughness modelling}.
\bjtitle{J. Fluid Mech.}
\bvolume{999},
\bfpage{78}
(\byear{2024})
\end{barticle}
\endbibitem

%%% 60
\bibitem[\protect\citeauthoryear{Guo et~al.}{2025}]{guo2025deepseek}
\begin{botherref}
\oauthor{\bsnm{Guo}, \binits{D.}},
\oauthor{\bsnm{Yang}, \binits{D.}},
\oauthor{\bsnm{Zhang}, \binits{H.}},
\oauthor{\bsnm{Song}, \binits{J.}},
\oauthor{\bsnm{Zhang}, \binits{R.}},
\oauthor{\bsnm{Xu}, \binits{R.}},
\oauthor{\bsnm{Zhu}, \binits{Q.}},
\oauthor{\bsnm{Ma}, \binits{S.}},
\oauthor{\bsnm{Wang}, \binits{P.}},
\oauthor{\bsnm{Bi}, \binits{X.}}, et al.:
Deepseek-{R1}: Incentivizing reasoning capability in {LLM}s via reinforcement learning.
arXiv preprint arXiv:2501.12948
(2025)
\end{botherref}
\endbibitem

%%% 61
\bibitem[\protect\citeauthoryear{Vreman}{2004}]{vreman2004eddy}
\begin{barticle}
\bauthor{\bsnm{Vreman}, \binits{A.}}:
\batitle{An eddy-viscosity subgrid-scale model for turbulent shear flow: Algebraic theory and applications}.
\bjtitle{Phys. Fluids}
\bvolume{16}(\bissue{10}),
\bfpage{3670}--\blpage{3681}
(\byear{2004})
\end{barticle}
\endbibitem

%%% 62
\bibitem[\protect\citeauthoryear{Piomelli et~al.}{1989}]{piomelli1989new}
\begin{barticle}
\bauthor{\bsnm{Piomelli}, \binits{U.}},
\bauthor{\bsnm{Ferziger}, \binits{J.}},
\bauthor{\bsnm{Moin}, \binits{P.}},
\bauthor{\bsnm{Kim}, \binits{J.}}:
\batitle{New approximate boundary conditions for large eddy simulations of wall-bounded flows}.
\bjtitle{Phys. Fluids}
\bvolume{1}(\bissue{6}),
\bfpage{1061}--\blpage{1068}
(\byear{1989})
\end{barticle}
\endbibitem

%%% 63
\bibitem[\protect\citeauthoryear{Pope}{2001}]{pope2001turbulent}
\begin{bbook}
\bauthor{\bsnm{Pope}, \binits{S.B.}}:
\bbtitle{Turbulent Flows}.
\bpublisher{Cambridge University Press},
\blocation{Cambridge, UK}
(\byear{2001})
\end{bbook}
\endbibitem

%%% 64
\bibitem[\protect\citeauthoryear{Xia et~al.}{2016}]{xia2016direct}
\begin{barticle}
\bauthor{\bsnm{Xia}, \binits{Z.}},
\bauthor{\bsnm{Shi}, \binits{Y.}},
\bauthor{\bsnm{Chen}, \binits{S.}}:
\batitle{Direct numerical simulation of turbulent channel flow with spanwise rotation}.
\bjtitle{J. Fluid Mech.}
\bvolume{788},
\bfpage{42}--\blpage{56}
(\byear{2016})
\end{barticle}
\endbibitem

%%% 65
\bibitem[\protect\citeauthoryear{Flack et~al.}{2016}]{flack2016skin}
\begin{barticle}
\bauthor{\bsnm{Flack}, \binits{K.A.}},
\bauthor{\bsnm{Schultz}, \binits{M.P.}},
\bauthor{\bsnm{Barros}, \binits{J.M.}},
\bauthor{\bsnm{Kim}, \binits{Y.C.}}:
\batitle{Skin-friction behavior in the transitionally-rough regime}.
\bjtitle{Int J Heat Fluid Flow}
\bvolume{61},
\bfpage{21}--\blpage{30}
(\byear{2016})
\end{barticle}
\endbibitem

%%% 66
\bibitem[\protect\citeauthoryear{Medjnoun et~al.}{2021}]{medjnoun2021turbulent}
\begin{barticle}
\bauthor{\bsnm{Medjnoun}, \binits{T.}},
\bauthor{\bsnm{Rodriguez-Lopez}, \binits{E.}},
\bauthor{\bsnm{Ferreira}, \binits{M.}},
\bauthor{\bsnm{Griffiths}, \binits{T.}},
\bauthor{\bsnm{Meyers}, \binits{J.}},
\bauthor{\bsnm{Ganapathisubramani}, \binits{B.}}:
\batitle{Turbulent boundary-layer flow over regular multiscale roughness}.
\bjtitle{J. Fluid Mech.}
\bvolume{917},
\bfpage{1}
(\byear{2021})
\end{barticle}
\endbibitem

%%% 67
\bibitem[\protect\citeauthoryear{Womack et~al.}{2022}]{womack2022turbulent}
\begin{barticle}
\bauthor{\bsnm{Womack}, \binits{K.M.}},
\bauthor{\bsnm{Volino}, \binits{R.J.}},
\bauthor{\bsnm{Meneveau}, \binits{C.}},
\bauthor{\bsnm{Schultz}, \binits{M.P.}}:
\batitle{Turbulent boundary layer flow over regularly and irregularly arranged truncated cone surfaces}.
\bjtitle{J. Fluid Mech.}
\bvolume{933},
\bfpage{38}
(\byear{2022})
\end{barticle}
\endbibitem

%%% 68
\bibitem[\protect\citeauthoryear{Flack and Schultz}{2023}]{flack2023hydraulic}
\begin{barticle}
\bauthor{\bsnm{Flack}, \binits{K.A.}},
\bauthor{\bsnm{Schultz}, \binits{M.P.}}:
\batitle{Hydraulic characterization of sandpaper roughness}.
\bjtitle{Exp. Fluids.}
\bvolume{64}(\bissue{1}),
\bfpage{3}
(\byear{2023})
\end{barticle}
\endbibitem

%%% 69
\bibitem[\protect\citeauthoryear{Yang et~al.}{2020}]{yang2020scaling}
\begin{barticle}
\bauthor{\bsnm{Yang}, \binits{X.I.A.}},
\bauthor{\bsnm{Pirozzoli}, \binits{S.}},
\bauthor{\bsnm{Abkar}, \binits{M.}}:
\batitle{Scaling of velocity fluctuations in statistically unstable boundary-layer flows}.
\bjtitle{J. Fluid Mech.}
\bvolume{886},
\bfpage{3}
(\byear{2020})
\end{barticle}
\endbibitem

%%% 70
\bibitem[\protect\citeauthoryear{Anderson et~al.}{2018}]{anderson2018turbulent}
\begin{barticle}
\bauthor{\bsnm{Anderson}, \binits{W.}},
\bauthor{\bsnm{Yang}, \binits{J.}},
\bauthor{\bsnm{Shrestha}, \binits{K.}},
\bauthor{\bsnm{Awasthi}, \binits{A.}}:
\batitle{Turbulent secondary flows in wall turbulence: vortex forcing, scaling arguments, and similarity solution}.
\bjtitle{Environmental Fluid Mechanics}
\bvolume{18},
\bfpage{1351}--\blpage{1378}
(\byear{2018})
\end{barticle}
\endbibitem

%%% 71
\bibitem[\protect\citeauthoryear{Abkar et~al.}{2016}]{abkar2016minimum}
\begin{barticle}
\bauthor{\bsnm{Abkar}, \binits{M.}},
\bauthor{\bsnm{Bae}, \binits{H.J.}},
\bauthor{\bsnm{Moin}, \binits{P.}}:
\batitle{Minimum-dissipation scalar transport model for large-eddy simulation of turbulent flows}.
\bjtitle{Phys. Rev. Fluids}
\bvolume{1}(\bissue{4}),
\bfpage{041701}
(\byear{2016})
\end{barticle}
\endbibitem

%%% 72
\bibitem[\protect\citeauthoryear{Martinez-Tossas et~al.}{2018}]{martinez2018comparison}
\begin{botherref}
\oauthor{\bsnm{Martinez-Tossas}, \binits{L.A.}},
\oauthor{\bsnm{Churchfield}, \binits{M.J.}},
\oauthor{\bsnm{Yilmaz}, \binits{A.E.}},
\oauthor{\bsnm{Sarlak}, \binits{H.}},
\oauthor{\bsnm{Johnson}, \binits{P.L.}},
\oauthor{\bsnm{S{\o}rensen}, \binits{J.N.}},
\oauthor{\bsnm{Meyers}, \binits{J.}},
\oauthor{\bsnm{Meneveau}, \binits{C.}}:
Comparison of four large-eddy simulation research codes and effects of model coefficient and inflow turbulence in actuator-line-based wind turbine modeling.
J. Renew. Sustain. Energy
\textbf{10}(3)
(2018)
\end{botherref}
\endbibitem

%%% 73
\bibitem[\protect\citeauthoryear{Yang et~al.}{2024}]{yang2024grid}
\begin{botherref}
\oauthor{\bsnm{Yang}, \binits{X.I.A.}},
\oauthor{\bsnm{Abkar}, \binits{M.}},
\oauthor{\bsnm{Park}, \binits{G.}}:
Grid convergence properties of wall-modeled large eddy simulations in the asymptotic regime.
J. Fluids Eng.
\textbf{146}(8)
(2024)
\end{botherref}
\endbibitem

%%% 74
\bibitem[\protect\citeauthoryear{Lv et~al.}{2021}]{lv2021wall}
\begin{botherref}
\oauthor{\bsnm{Lv}, \binits{Y.}},
\oauthor{\bsnm{Huang}, \binits{X.L.}},
\oauthor{\bsnm{Yang}, \binits{X.}},
\oauthor{\bsnm{Yang}, \binits{X.I.A.}}:
Wall-model integrated computational framework for large-eddy simulations of wall-bounded flows.
Phys. Fluids
\textbf{33}(12)
(2021)
\end{botherref}
\endbibitem

%%% 75
\bibitem[\protect\citeauthoryear{Gao and Lv}{2025}]{gao2025novel}
\begin{botherref}
\oauthor{\bsnm{Gao}, \binits{R.}},
\oauthor{\bsnm{Lv}, \binits{Y.}}:
A novel large-eddy simulation framework with consistently enforced wall models by a near-wall dynamic correction procedure.
Phys. Fluids
\textbf{37}(3)
(2025)
\end{botherref}
\endbibitem

%%% 76
\bibitem[\protect\citeauthoryear{Bobke et~al.}{2017}]{bobke2017history}
\begin{barticle}
\bauthor{\bsnm{Bobke}, \binits{A.}},
\bauthor{\bsnm{Vinuesa}, \binits{R.}},
\bauthor{\bsnm{{\"O}rl{\"u}}, \binits{R.}},
\bauthor{\bsnm{Schlatter}, \binits{P.}}:
\batitle{History effects and near equilibrium in adverse-pressure-gradient turbulent boundary layers}.
\bjtitle{J. Fluid Mech.}
\bvolume{820},
\bfpage{667}--\blpage{692}
(\byear{2017})
\end{barticle}
\endbibitem

%%% 77
\bibitem[\protect\citeauthoryear{Perry et~al.}{2002}]{perry2002streamwise}
\begin{barticle}
\bauthor{\bsnm{Perry}, \binits{A.}},
\bauthor{\bsnm{Marusic}, \binits{I.}},
\bauthor{\bsnm{Jones}, \binits{M.}}:
\batitle{On the streamwise evolution of turbulent boundary layers in arbitrary pressure gradients}.
\bjtitle{J. Fluid Mech.}
\bvolume{461},
\bfpage{61}--\blpage{91}
(\byear{2002})
\end{barticle}
\endbibitem

%%% 78
\bibitem[\protect\citeauthoryear{Volino}{2020}]{volino2020reynolds}
\begin{barticle}
\bauthor{\bsnm{Volino}, \binits{R.J.}}:
\batitle{{R}eynolds number dependence of zero pressure gradient turbulent boundary layers including third-order moments and spatial correlations}.
\bjtitle{J. Fluids Eng.}
\bvolume{142}(\bissue{5}),
\bfpage{051303}
(\byear{2020})
\end{barticle}
\endbibitem

%%% 79
\bibitem[\protect\citeauthoryear{Pozuelo et~al.}{2022}]{pozuelo2022adverse}
\begin{barticle}
\bauthor{\bsnm{Pozuelo}, \binits{R.}},
\bauthor{\bsnm{Li}, \binits{Q.}},
\bauthor{\bsnm{Schlatter}, \binits{P.}},
\bauthor{\bsnm{Vinuesa}, \binits{R.}}:
\batitle{An adverse-pressure-gradient turbulent boundary layer with nearly constant}.
\bjtitle{J. Fluid Mech.}
\bvolume{939},
\bfpage{34}
(\byear{2022})
\end{barticle}
\endbibitem

%%% 80
\bibitem[\protect\citeauthoryear{Johnston et~al.}{1972}]{johnston1972effects}
\begin{barticle}
\bauthor{\bsnm{Johnston}, \binits{J.P.}},
\bauthor{\bsnm{Halleent}, \binits{R.M.}},
\bauthor{\bsnm{Lezius}, \binits{D.K.}}:
\batitle{Effects of spanwise rotation on the structure of two-dimensional fully developed turbulent channel flow}.
\bjtitle{J. Fluid Mech.}
\bvolume{56}(\bissue{3}),
\bfpage{533}--\blpage{557}
(\byear{1972})
\end{barticle}
\endbibitem

%%% 81
\bibitem[\protect\citeauthoryear{Brethouwer}{2017}]{brethouwer2017statistics}
\begin{barticle}
\bauthor{\bsnm{Brethouwer}, \binits{G.}}:
\batitle{Statistics and structure of spanwise rotating turbulent channel flow at moderate {R}eynolds numbers}.
\bjtitle{J. Fluid Mech.}
\bvolume{828},
\bfpage{424}--\blpage{458}
(\byear{2017})
\end{barticle}
\endbibitem

%%% 82
\bibitem[\protect\citeauthoryear{Barlow and Coceal}{2008}]{barlow2008review}
\begin{botherref}
\oauthor{\bsnm{Barlow}, \binits{J.}},
\oauthor{\bsnm{Coceal}, \binits{O.}}:
A review of urban roughness sublayer turbulence.
Technical report,
Met Office
(2008)
\end{botherref}
\endbibitem

%%% 83
\bibitem[\protect\citeauthoryear{Bons}{2010}]{bons2010review}
\begin{barticle}
\bauthor{\bsnm{Bons}, \binits{J.P.}}:
\batitle{A review of surface roughness effects in gas turbines}.
\bjtitle{J. Turbomach.}
\bvolume{132}(\bissue{2}),
\bfpage{021004}
(\byear{2010})
\end{barticle}
\endbibitem

%%% 84
\bibitem[\protect\citeauthoryear{Schultz}{2007}]{schultz2007effects}
\begin{barticle}
\bauthor{\bsnm{Schultz}, \binits{M.P.}}:
\batitle{Effects of coating roughness and biofouling on ship resistance and powering}.
\bjtitle{Biofouling}
\bvolume{23}(\bissue{5}),
\bfpage{331}--\blpage{341}
(\byear{2007})
\end{barticle}
\endbibitem

%%% 85
\bibitem[\protect\citeauthoryear{Chung et~al.}{2021}]{chung2021predicting}
\begin{barticle}
\bauthor{\bsnm{Chung}, \binits{D.}},
\bauthor{\bsnm{Hutchins}, \binits{N.}},
\bauthor{\bsnm{Schultz}, \binits{M.P.}},
\bauthor{\bsnm{Flack}, \binits{K.A.}}:
\batitle{Predicting the drag of rough surfaces}.
\bjtitle{Annu. Rev. Fluid. Mech.}
\bvolume{53}(\bissue{1}),
\bfpage{439}--\blpage{471}
(\byear{2021})
\end{barticle}
\endbibitem

%%% 86
\bibitem[\protect\citeauthoryear{Nikuradse}{1950}]{nikuradse1950laws}
\begin{botherref}
\oauthor{\bsnm{Nikuradse}, \binits{J.}}:
Laws of flow in rough pipes.
Technical Report Technical Memorandum 1292,
National Advisory Committee for Aeronautics
(1950)
\end{botherref}
\endbibitem

%%% 87
\bibitem[\protect\citeauthoryear{Colebrook}{1939}]{colebrook1939correspondence}
\begin{barticle}
\bauthor{\bsnm{Colebrook}, \binits{C.F.}}:
\batitle{Turbulent flow in pipes, with particular reference to the transition region between the smooth and rough pipe laws}.
\bjtitle{J. Inst. Civ. Eng.}
\bvolume{11}(\bissue{4}),
\bfpage{133}--\blpage{156}
(\byear{1939})
\end{barticle}
\endbibitem

%%% 88
\bibitem[\protect\citeauthoryear{Moody}{1944}]{moody1944friction}
\begin{barticle}
\bauthor{\bsnm{Moody}, \binits{L.F.}}:
\batitle{Friction factors for pipe flow}.
\bjtitle{Trans. ASME.}
\bvolume{66}(\bissue{8}),
\bfpage{671}--\blpage{678}
(\byear{1944})
\end{barticle}
\endbibitem

%%% 89
\bibitem[\protect\citeauthoryear{Flack and Schultz}{2014}]{flack2014roughness}
\begin{botherref}
\oauthor{\bsnm{Flack}, \binits{K.A.}},
\oauthor{\bsnm{Schultz}, \binits{M.P.}}:
Roughness effects on wall-bounded turbulent flows.
Phys. Fluids
\textbf{26}(10)
(2014)
\end{botherref}
\endbibitem

%%% 90
\bibitem[\protect\citeauthoryear{Forooghi et~al.}{2017}]{forooghi2017toward}
\begin{barticle}
\bauthor{\bsnm{Forooghi}, \binits{P.}},
\bauthor{\bsnm{Stroh}, \binits{A.}},
\bauthor{\bsnm{Magagnato}, \binits{F.}},
\bauthor{\bsnm{Jakirli{\'c}}, \binits{S.}},
\bauthor{\bsnm{Frohnapfel}, \binits{B.}}:
\batitle{Toward a universal roughness correlation}.
\bjtitle{J. Fluids Eng.}
\bvolume{139}(\bissue{12}),
\bfpage{121201}
(\byear{2017})
\end{barticle}
\endbibitem

%%% 91
\bibitem[\protect\citeauthoryear{Rezwana and Maher}{2022}]{rezwana2022identifying}
\begin{botherref}
\oauthor{\bsnm{Rezwana}, \binits{J.}},
\oauthor{\bsnm{Maher}, \binits{M.L.}}:
Identifying ethical issues in ai partners in human-{AI} co-creation.
arXiv preprint arXiv:2204.07644
(2022)
\end{botherref}
\endbibitem

\end{thebibliography}
%% if required, the content of .bbl file can be included here once bbl is generated
%%\input sn-article.bbl

\end{document}